\documentclass[letterpaper, 11pt]{article}
\pdfoutput = 1

\usepackage{url}
\usepackage{setspace} 
\usepackage[bookmarks = false, colorlinks = true, linkcolor = blue, citecolor = purple]{hyperref}     
\usepackage{shortcuts}

\usepackage{tikz}
\usetikzlibrary{arrows,decorations.pathmorphing,backgrounds,positioning,fit,petri}

\usepackage[margin = 2.5cm]{geometry} 
\titleformat*{\section}{\Large \bfseries}
\titleformat*{\subsection}{\large \bfseries }
\titleformat*{\subsubsection}{\normalsize \itshape}

\begin{document}

\begin{center}

\thispagestyle{empty}

\vspace*{5em}

{\LARGE \bf Comments on Defining Entanglement Entropy}

\vspace{1cm}

{\large Jennifer Lin$^\spadesuit$ and \DJ or\dj e Radi\v cevi\'c$^\clubsuit$}
\vspace{1em}

$^\spadesuit${\it School of Natural Sciences, Institute for Advanced Study, Princeton, NJ, USA}\\
\texttt{jenlin@ias.edu}
\\
\vspace{1.5em}

$^\clubsuit${\it Perimeter Institute for Theoretical Physics, Waterloo, Ontario, Canada N2L 2Y5}\\
\texttt{djordje@pitp.ca}\\

\vspace{0.08\textheight}
\begin{abstract}
We revisit the issue of defining the entropy of a spatial region in a broad class of quantum theories. In theories with explicit regularizations, working within an elementary but general algebraic framework applicable to matter and gauge theories alike, we give precise path integral expressions for three known types of entanglement entropy that we call full, distillable, and gauge-invariant. For a class of gauge theories that do not necessarily have a regularization in our framework, including Chern-Simons theory, we describe a related approach to defining entropies based on locally extending the Hilbert space at the entangling edge, and we discuss its connections to other calculational prescriptions. Based on results from both approaches, we conjecture that it is always the full entanglement entropy that is calculated by standard holographic techniques in strongly coupled conformal theories.
\end{abstract}
\end{center}

\newpage
\tableofcontents

\section{Introduction and summary}

Studying the structure of entanglement in quantum field theory and many-body systems has long been the focus of a tremendous amount of activity.\footnote{See, for instance, \cite{Srednicki:1993im,  Vidal:2002rm, Calabrese:2004eu, Kitaev:2005dm, Ryu:2006bv, Li:2008, VanRaamsdonk:2010pw} for a very incomplete sampling of important insights; see \cite{Amico:2007ag} for a venerable review, \cite{Nishioka:2018khk} for a modern QFT-oriented review, and \cite{Preskill:1999he} for a succinct and prescient summary of many relevant ideas.} One of the most fundamental measures of entanglement in such systems is the entropy associated to a spatial subregion. This paper is fully devoted to exploring certain subtle issues that arise in defining this quantity.

It is common --- but imprecise --- to define the entanglement entropy $S_{\bb V}$ of a spatial region $\bb V$ in a quantum state $\rho$ as the von Neumann entropy, $-\Tr \rho_{\bb V}\log \rho_{\bb V}$, of the reduced density matrix $\rho_{\bb V}$ obtained by tracing out the degrees of freedom outside of $\bb V$, i.e.~as $\rho_{\bb V} = \Tr_{\bar{\bb V}} \rho$. The issue is that this definition assumes that the Hilbert space factorizes into a direct product of degrees of freedom on $\bb V$ and its complement $\bar{\bb V}$. This assumption fails in many cases, including in strictly continuum QFTs (see the remarkably clear note \cite{Witten:2018zxz} and references within) and in any lattice gauge theory \cite{Buividovich:2008gq, Donnelly:2011hn, Casini:2013rba, Radicevic:2014kqa, Donnelly:2014gva, Ghosh:2015iwa, Soni:2015yga}.

A more precise approach is to define $S_{\bb V}$ as the entropy of an algebra of operators $\A_{\bb V}$ supported on $\bb V$ (see \cite{Zanardi:2004zz, ohya2004quantum} and references therein). If the Hilbert space does factorize and if $\A_{\bb V}$ is the \emph{maximal} algebra supported on $\bb V$, its entropy agrees with the one defined via tracing out. If the Hilbert space does not factorize but the theory is fully regularized (e.g.~it is defined on a finite lattice), the algebraic definition of entropy is meaningful even though tracing out is not defined. If the theory is defined directly in the continuum, the entropy associated to a subalgebra is not definable --- but related quantities, such as relative entropy, will still be well-defined algebraically.

A catch in the above definition is that there is no unique algebra supported on a given region. For instance, even if the Hilbert space factorizes, if $\A_{\bb V}$ is not the maximal algebra on $\bb V$, the associated entropy will not generically agree with the tracing-out prescription. Moreover, for every non-maximal algebra there will be several natural entropic measures that can be associated to it. This is well known and has been discussed in, for example, refs.~\cite{Casini:2013rba, Soni:2016ogt}.

In this paper we will explore these algebraic choices in more generality. We will focus on theories with explicit regularizations: spin systems, fermion and scalar lattice field theories, and lattice gauge theories. (It is perhaps worth stressing that all of these theories require similar choices in their definitions of entropies; the issues we explore are \emph{not} fundamentally due to the non-factorizability of gauge theory Hilbert spaces.) Our goal is to clarify \emph{which} entanglement entropy one talks about when computing it using various prescriptions that exist in the literature. Many of the ideas here are not new; here we publicize them further and show how they are connected to each other.

Our main focus will be on replica trick path integrals, which are perhaps the most useful way of computing entanglement entropy in field theories. One of our main results is an explicit dictionary that, for a fixed entangling region $\bb V$, translates between certain choices of algebras and boundary conditions in these path integrals. (Basic aspects of this connection were presented in \cite{Jafferis:2015del}.) We will identify algebras that correspond to open, Dirichlet, and Neumann boundary conditions on the entanglement edge $\del \bb V$ in path integrals with matter fields. In gauge theories, we will similarly identify algebras corresponding to natural gauge-invariant boundary conditions that we will call electric and magnetic. All of these boundary conditions are imposed on the entanglement edge only on a \emph{single} time slice, and at other times the fields are unconstrained.

We will also demonstrate several new points. One is that when a particular choice of algebra $\A_{\bb V}$ corresponds to a particular type of a boundary condition, then it generically corresponds to a combination of \emph{all} possible conditions of that type. For instance, an algebra corresponding to Dirichlet conditions for a scalar field $\varphi(x)$ will have a natural entropy calculated by a (weighted) sum over all possible boundary conditions $\{k(x_\parallel)\}$ of the type $\lim_{x \rar x_\parallel} \varphi(x) = k(x_\parallel)$ for $x_\parallel \in \del\bb V$. Another important lesson will be that the choices of algebras have relatively little to do with the entangling edge itself. While some choices of $\A_{\bb V}$ will correspond to boundary conditions of various sorts on $\del \bb V$, other choices will correspond to ``boundary'' conditions (or constraints) in the interior of $\bb V$. The choices in the definition of entanglement entropy that we discuss here are thus much more general than the ambiguities in introducing a regulator for the conical defect in the replica path integral, and our analysis holds even in these more general cases.

The analysis described so far, being rather microscopic and micromanaging in nature, will not apply to theories that do not have known lattice discretizations. Examples include many field theories of interest, including chiral theories in even spacetime dimensions and Chern-Simons theory. However, a different prescription for calculating entanglement entropy allows us to say a bit about some of these theories, too \cite{Levin:2006zz, Buividovich:2008gq, Donnelly:2011hn, Ghosh:2015iwa, Soni:2015yga, Pretko:2015zva}. The procedure in question involves local extensions of the Hilbert space that make it factorize on the entanglement edge: instead of regularizing these theories fully, we add the minimal amount of degrees of freedom needed to regularize the entangling edge alone. This allows us to apply the original tracing-out prescription to e.g.~Chern-Simons theory, and to draw some instructive parallels to the algebraic approach.

Having thus whetted the reader's appetite, we overview the structure of the paper. In section \ref{sec formal}, we set the notation and carefully review some dry (but important) facts about finite systems. We set up the analysis so that it works for \emph{arbitrary} lattice theories, so our results will have obvious analogs in continuum scalar, matter, and gauge theories with known actions and field content.

In section \ref{sec algebras}, we connect different operator algebras on spatial lattices $\bb M$ to different types of boundary conditions on boundaries $\del \bb M$ of these lattices. This is a warm-up: we focus on pure states, and there are no entangling regions, reduced density matrices, replica tricks, etc. This analysis is also important unto itself, e.g.~for the purposes of analyzing exact dualities of quantum theories on manifolds with boundaries.

In section \ref{sec subalgebras} we perform a similar analysis for algebras on subregions $\bb V \subset \bb M$. The only difference between scalar and gauge theories arises at this step: for scalar theories, maximal subalgebras will never have a center, while for gauge theories, maximal gauge-invariant subalgebras always will. We will establish how analyzing the center of a chosen algebra $\A_{\bb V}$ tells us all we need to know about the type of boundary condition that will be obeyed by the appropriate reduced density matrix.

In section \ref{sec entropies} we introduce several measures of entanglement that can be associated to the reduced density matrices constructed so far. In section \ref{sec path integrals} we show how these entropies are computed using replica trick path integrals, and we demonstrate how choices of algebras are reflected on the path integral side. In these sections we will focus on three natural types of entropies: the \emph{full} entropy, the \emph{distillable} entropy, and the \emph{gauge-invariant} entropy. (These will be three out of many possible entropies one can define.) We will describe how they are related to each other and give precise path integral expressions for each of them.

In section \ref{sec comments} we present several important remarks that that connect our analysis to other approaches of calculating the entanglement entropy. For instance, we comment on how choices of algebras in conformal field theories are reflected by conformal boundary conditions in the 2D Ising model, based on the analysis in \cite{Ohmori:2014eia}. In the context of CFTs more generally, we also argue that the full and gauge-invariant entanglement entropies are natural candidates for the types of entropies that may contain universal information on trace anomalies, meaning that either \emph{could} agree with the holographic entanglement entropy of Ryu and Takayanagi \cite{Ryu:2006bv}.

In section \ref{s3} we return to the discussion of entropies in theories where a lattice realization is unknown, where we employ local extensions of the Hilbert space. We review a cross-section of the existing literature on entanglement entropy in gauge theories, present some archetypical calculations using the extended Hilbert space, and suggest how they fit into our general paradigm. Building on this, we extend our earlier arguments and conjecture that it is precisely the \emph{full} entanglement entropy that contains the universal CFT information obtained using holography and other techniques. We conclude and stress a few open problems in section \ref{sec conclusion}.

\section{Formal preliminaries} \label{sec formal}

Consider a Hilbert space $\H$ composed of $N$ copies of a $K$-dimensional \emph{target space} $\H_0$,
\bel{\label{def H}
  \H = \bigotimes_{i = 1}^N \H_i \simeq \H_0^{\otimes N}, \quad D \equiv \dim \H = K^N.
}
The use of a regular direct product means that this system is bosonic. For fermionic theories we should use graded products. Instead, we exploit the fact that fermionic systems can be dualized to bosonic systems in which the target has dimension $K = 2$; this is possible even in higher dimensions, with only mild assumptions on the regularity of spatial lattices \cite{Chen:2017fvr, Chen:2018nog}. This way we can always work with a Hilbert space $\H$ defined as a conventional direct product.

The maximal algebra of operators that acts on this space is the algebra of complex matrices $\C^{D \times D}$. Recall that an algebra is a vector space (in this case, over $\C$) equipped with and closed under a vector product (in this case, matrix multiplication). This maximal algebra has $D^2$ complex dimensions. It is very convenient to express the basis as the set of all possible products of a small number of generators. Due to the direct product structure of $\H$, it is natural to pick the basis of $\C^{D \times D}$ to be generated by operators of the form
\bel{\label{op dir prod}
  \1_1 \otimes \1_2 \otimes \ldots \otimes \1_{i - 1} \otimes \O_i \otimes \1_{i + 1} \otimes \ldots \otimes \1_N,
}
where $\O_i$ is an operator that generates a basis of $\C^{K \times K}$, the maximal algebra on the $i$'th target space. For brevity, we will drop the $\otimes$ signs and factors of $\1$, so $\O_i$ will denote the entire direct product \eqref{op dir prod}.

Only two generators are needed to generate the full basis of $\H_i$, and we will call them the generalized position and momentum generators $\Phi_i$ and $\Pi_i$ at location $i$. One simple choice is
\bel{\label{def Phi Pi}
  \Phi_i = \left[
             \begin{array}{cccc}
               1 &  &  &  \\
                & \e^{2\pi \i / K} &  &  \\
                &  & \ddots &  \\
                &  &  & \e^{2\pi \i (K - 1)/K} \\
             \end{array}
           \right],
  \quad
  \Pi_i = \left[
            \begin{array}{cccc}
                &  &  & 1 \\
               1&  &  &  \\
                & 1 &  &  \\
                &  & \ddots &  \\
            \end{array}
          \right].
}
For a clock model (a discretized version of the compact scalar theory) these are actual position and momentum operators. A particular Hamiltonian may describe theories very different from this one, e.g.~a $\sigma$-model whose target space is a (discretization of) a curved manifold, but any such theory will have an operator algebra isomorphic to the one generated by the above $\Phi_i$ and $\Pi_i$.

From now on, ``generators of an algebra'' will refer to a set of operators $\G$ whose all possible products form a basis that spans the entire algebra. For a general case of $N$ sites with a $K$-dimensional target space on each site, the canonical choice for generators of $\C^{D \times D}$ is $\G = \{\Phi_i, \Pi_i\}_{i = 1, \ldots, N}$. For a spin system, the target space has $K = 2$ and the canonical generators from eq.~\eqref{def Phi Pi} are the Pauli matrices, $\Phi_i = Z_i$ and $\Pi_i = X_i$. For spinless complex fermions, one typically chooses $\G\_{ferm} = \{\psi_i, \psi_i\+\}_{i = 1, \ldots, N}$, with $\{\psi_i,\psi\+_i\} \propto \1$ and $\{\psi_i, \psi_j\} = \{\psi_i, \psi_j\+\} = 0$. The algebra generated by $\G\_{ferm}$ is isomorphic to that of a $K = 2$ bosonic system.\footnote{This isomorphism can be made very explicit in one spatial dimension, where it is known as the Jordan-Wigner transformation. It maps a chain of fermions $\psi_v$ for $v = 1,\ldots, N$ to a spin chain, via $\psi_v = (X_v + \i Y_v) \prod_{u = 1}^{v - 1} Z_u$. }

A pure state $\qvec\psi \in \H$ induces expectation values $\avg \O = \qmat\psi\O\psi$ of all operators in $\C^{D \times D}$. More generally, a density matrix is a Hermitian operator $\rho \in \C^{D \times D}$ that induces expectation values via
\bel{
  \avg\O = \Tr(\rho\O).
}
Conversely, any set of expectation values induces a density matrix. To see this, it is very useful to choose an orthonormal basis $\{\O_a\}_{a = 1,\ldots, D^2}$ for $\C^{D \times D}$ where all $\O_a$ are invertible and, except the identity, traceless. The orthonormality is defined with respect to the natural trace inner product, such that $\Tr(\O_a^{-1} \O_b) = D\,\delta_{ab}$ for all $a$, $b$. One example is the canonical basis generated by $\Phi_i$ and $\Pi_i$ in eq.~\eqref{def Phi Pi}. The explicit expansion of the density matrix in any such basis is
\bel{\label{def rho}
  \rho = \frac 1 D \sum_{a = 1}^{D^2} \avg{\O_a^{-1}} \O_a.
}

Let us now consider non-maximal algebras $\A \subset \C^{D \times D}$. For our purposes, the relevant objects are unital \hbox{$*$-algebras} $\A$. These are algebras that contain the identity and that are closed under Hermitian conjugation.\footnote{An algebra encodes the allowed operations that can be performed on a given quantum system. We require that $\A$ be unital because we should always be allowed to leave the system as it is, by acting on it with the identity. We require that $\A$ be a $*$-algebra because $\rho$ is necessarily Hermitian, so operators $\O$ that appear in the expansion  \eqref{def rho} of $\rho$ must either be Hermitian themselves, or must come in pairs of Hermitian conjugates. If an operator is in $\A$ but its conjugate is not, then it cannot appear in  \eqref{def rho}. Any density matrix in $\A$ thus also belongs to the maximal subalgebra of $\A$ that is a $*$-algebra. Therefore the entire discussion may be phrased from the outset for $*$-algebras alone.} In practice, we will only look at non-maximal algebras generated by different products of generalized positions and momenta $\Phi_i$ and $\Pi_i$, e.g.~$\{\Phi_i \Phi_{i + 1}, \Pi_i\}_{i = 1, \ldots, N}$ or $\{\Phi_i, \Pi_i\}_{i = 1, \ldots, M}$ for $M < N$. Not only will algebras thus obtained all be unital and $*$, the bases generated by these generators will also automatically be orthonormal, with all basis operators traceless (except the identity) and invertible.

Given a non-maximal, unital $*$-algebra $\A$ spanned by a basis $\{\O_{a'}\}$ of the kind just described, there exists a unique density matrix $\rho \in \A$ that reproduces any list of specified expectations of elements of $\A$ via $\rho = \frac1D \sum_{a'} \avg{\O^{-1}_{a'}} \O_{a'}$. At this level of generality, $\rho$ is still a $D\times D$ matrix, i.e.~an operator acting on the full Hilbert space $\H$.

An important fact is that any operator that is \emph{not} in $\A$ and has zero inner product with all operators in $\A$ will have a vanishing expectation in state $\rho$. This is an abstract form of Elitzur's theorem \cite{Elitzur:1975im}. A reduction of the algebra from $\C^{D \times D}$ to $\A$ is intimately related to gauging degrees of freedom, i.e.~to imposing constraints on the space of allowed density matrices.

As a concrete and nontrivial example of these ideas, let us explore the connection between non-maximal algebras and gauging in the context of pure $\Z_2$ gauge theories in $d = 2$ spatial dimensions.\footnote{Henceforth we always use the lowercase $d$ for the number of spatial dimensions.} This will also set the stage for our discussion of gauge theories in general. Here we start with the full (ungauged) Hilbert space $\H$, given as in eq.~\eqref{def H} by a tensor product of a $K = 2$ bosonic Hilbert space over each link $\ell$ of a lattice $\Mbb$. This is the setup of Kitaev's toric code \cite{Kitaev:1997wr}: no gauge constraint has been introduced yet.\footnote{\label{f4}From a traditional gauge theory perspective, one could refer to $\H$ as a \emph{(globally) extended Hilbert space}, the understanding being that only the gauge-invariant states are physical and that $\H$ simply provides a convenient embedding for the space of gauge-invariant states. The distinction between gauge-invariant and full/globally extended Hilbert spaces does not influence any correlation functions of the gauge theory but does appear in calculations of entanglement entropy, leading us to distinguish between gauge-invariant and full entropies in section \ref{sec entropies}. In sections \ref{sec formal} through \ref{sec comments} we will use the globally extended Hilbert space whenever discussing lattice gauge theories, without further comment. In section \ref{s3} we will say more about this choice and define measures of entanglement referring to only the gauge-invariant Hilbert space. These entropies will be equivalent to what we call full and gauge-invariant entropies when an explicit regularization is available.} The maximal algebra of observables is canonically generated by Pauli matrices on links, $\{Z_\ell, X_\ell\}$. Now, consider the algebra $\A$ generated by $\{X_\ell, W_f\}$, where $W_f = \prod_{\ell \subset f} Z_\ell$ are products of $Z$ operators along faces (plaquettes) $f$.\footnote{On lattices of nontrivial topology, one may add to $\A$ gauge-invariant products of operators along noncontractible cycles. If we do not, $\A$ will contain a set of central generators associated to one-form symmetries. These operators are crucial for understanding dualities, for instance, but we will ignore them here.} This algebra has a huge center generated by Gauss operators $G_v = \prod_{\ell \supset v} X_\ell$, one per vertex $v$ of the lattice. Elements of $\A$ are precisely the familiar \emph{gauge-invariant} operators.

The fact that the gauge-invariant algebra has a center means that any density matrix $\rho \in \A$ must be block-diagonal in a Hilbert space basis that diagonalizes the center. In other words, the original Hilbert space splits into superselection sectors, one for each of the two possible eigenvalues $\pm 1$ of each $G_v$. For a connected lattice with $N$ vertices and no special boundary conditions, this means that there are $2^{N - 1}$ sectors (the product of $G_v$ over all sites is identically the identity, hence there is one central element fewer than would be naively expected). For every state whose density matrix is gauge-invariant, the expectation of any operator orthogonal to $\A$ --- say, $Z_\ell$ --- is zero. This is the original Elitzur's theorem.

In a $\Z_2$ gauge theory, the Gauss law $G_v = \1$ is usually imposed at each vertex. (Sometimes background sources are inserted, meaning that $G_v = -\1$ is imposed on a finite number of sites.) This operator equation is tantamount to restricting our attention to just one superselection sector of the full Hilbert space $\H$. (Note that Elitzur's theorem holds even without this restriction to a single sector.) In the toric code, and more generally in condensed matter gauge theories, these gauge constraints are imposed dynamically, by adding $-\sum_v G_v$ to the Hamiltonian and then focusing on the low-energy sector; in particle physics gauge theories, this constraint is imposed by fiat.

The gauge theory example is an instance of a more general phenomenon: a non-maximal algebra always has an associated center, and eigenvalues of generators of the center label different superselection sectors. All density matrices are direct sums (statistical mixtures) of smaller density matrices that act only within specific sectors. Sometimes, as with gauge-invariant algebras and states obeying the Gauss law in the preceding paragraph, only one superselection sector will be populated. At other times, as with more general subalgebras and states to be discussed below, multiple sectors will be populated. In those cases, within each sector, its labels can be interpreted as constraints. When the center generators have support only on the edges of a spatial lattice, these sector labels denote different boundary conditions and are called \emph{edge modes}.

\section{Algebras and boundary conditions} \label{sec algebras}

Consider a system defined on a spatial lattice $\Mbb$ with boundary.\footnote{The boundary of a 2d lattice is the set of all links that do not belong to exactly two faces. On a general triangulated $d$-manifold, the boundary is the set of all codimension-one simplices which do not belong to two codimension-zero simplices, together with any nonzero-codimension simplices that do not belong to any codimension-zero simplices.} There exist several families of algebras that are supported on all of $\Mbb$, but that differ in the choice of generators present at the boundary $\del \Mbb$. The purpose of this section is to show which families naturally correspond to which familiar sets of (lattice) boundary conditions.

As a warm-up, we start with a spin chain with $N$ sites. Here we will study families of algebras obtained by removing edge generators from the maximal algebra $\C^{D \times D}$ with $D = 2^N$. The generalization to higher dimensions and arbitrary compact scalar theories is immediate.\footnote{Noncompact scalars, on the other hand, do not have an algebraic formulation unless they are embedded into a compact scalar theory. We will only discuss compact target spaces in this paper.} Fermion systems in one spatial dimension will be examined next, and they also generalize to higher dimensions and parafermionic models. Finally, we will show how edge modes arise in gauge theories, where subalgebras of the non-maximal, gauge-invariant algebra $\A$ are considered.

\subsection{Spin chains and scalar matter}

A spin chain has a maximal algebra generated by Pauli operators $X_v$ and $Z_v$ on each site $v$. (An analogous discussion applies if we start from other generators, say $X_v$ and $Y_v$.) Consider now the algebra $\A\_{D,O}$ generated by
\bel{
  \G\_{D,O} = \{X_1, \ldots, X_N, Z_2, \ldots, Z_N\}.
}
The difference  from the maximal algebra is that one generator from the edge of the system, $Z_1$, is removed. The center of $\A\_{D,O}$ is generated by $X_1$, the remaining generator on the edge. All pure states with density matrices in this reduced algebra thus must be eigenstates of $X_1$. Moreover, a general density matrix will be a statistical mixture of $X_1 = \1$ and $X_1 = -\1$ states. Thus algebras missing one generator on the edge correspond to (mixtures of) states with definite values of spins at that edge. In the case of $\A\_{D,O}$, there are thus Dirichlet boundary conditions on one edge, and open boundary conditions on the other edge.

We can also consider ``coarse graining'' generators on two adjacent sites near the edge, so the remaining ones are only sensitive to some (but not all) degrees of freedom near $\del \Mbb$. For instance, take an algebra that cannot measure the $z$-component (magnetization) of individual spins at sites $1$ and $2$, but that can measure the parity of the \emph{total} magnetization on these sites. This algebra is generated by
\bel{
  \G\_{N,O} = \{X_1, \ldots, X_N, Z_1 Z_2, Z_3, Z_4, \ldots, Z_N\}.
}
The resulting algebra, $\A\_{N,O}$, has the same number of generators as the Dirichlet-open one, $\A\_{D,O}$. Its center is generated by $X_1 X_2$, and so states whose density matrices belong to this algebra are (statistical mixtures of) eigenstates of $X_1X_2$. If eigenvalues of $X_v$ are labeled by $\e^{\i \pi \phi_v}$ with $\phi_v \in \Z\, \trm{mod}\, 2$, then this algebra contains density matrices of states with definite values of $\phi_1 - \phi_2 \sim \del \phi_1$. This means that Neumann boundary conditions are imposed on one end of the chain, while the other end has open boundary conditions.

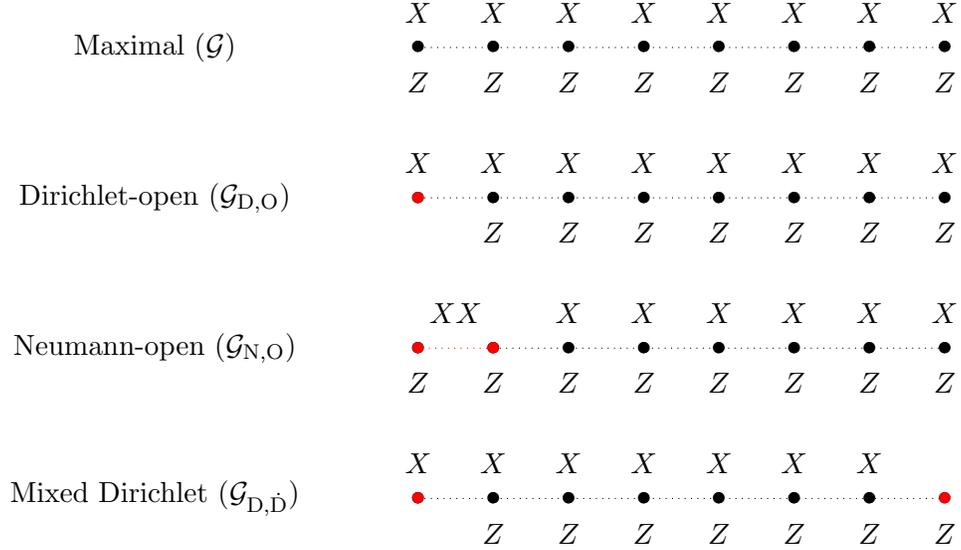
\begin{figure}[tb!]
\begin{center}

\begin{tikzpicture}[scale = 2]

  \draw[step = 0.5, dotted] (-1.75, 0) -- (1.75, 0);

  \foreach \x in {-1.75, -1.25, ..., 1.75}
    \draw (\x, 0) node {$\bullet$};



  \draw (-3.5, 0) node {Maximal ($\G$)};

  \foreach \x in {-1.75, -1.25, ..., 1.75} {
    \draw (\x, 0) node {$\bullet$};
    \draw (\x, -0.1) node[anchor = north] {$Z$};
    \draw (\x, 0.1) node[anchor = south] {$X$};
  };


  \draw[step = 0.5, dotted] (-1.75, -1) -- (1.75, -1);

  \foreach \x in {-1.75, -1.25, ..., 1.75}
    \draw (\x, -1) node {$\bullet$};


  \foreach \x in {-1.25, -0.75, ..., 1.75} {
    \draw (\x, -1.1) node[anchor = north] {$Z$};
  };

  \draw (-3.5, -1) node {Dirichlet-open ($\G\_{D, O}$)};

  \foreach \x in {-1.75, -1.25, ..., 1.75} {
    \draw (\x, -1) node {$\bullet$};
    \draw (\x, -0.9) node[anchor = south] {$X$};
  };

  \draw (-1.75, -1) node[red] {$\bullet$};


  \draw[step = 0.5, dotted] (-1.25, -2) -- (1.75, -2);

  \foreach \x in {-1.75, -1.25, ..., 1.75}
    \draw (\x, -2) node {$\bullet$};



  \draw (-3.5, -2) node {Neumann-open ($\G\_{N, O}$)};

  \foreach \x in {-0.75, -0.25, ..., 1.75} {
    \draw (\x, -1.9) node[anchor = south] {$X$};
  };

  \foreach \x in {-1.75, -1.25, ..., 1.75} {
    \draw (\x, -2) node {$\bullet$};
    \draw (\x, -2.1) node[anchor = north] {$Z$};
  };

  \draw (-1.5, -1.9) node[anchor = south] {$XX$};

  \draw[dotted, red] (-1.25, -2) -- (-1.75, -2);
  \draw (-1.25, -2) node[red] {$\bullet$};
  \draw (-1.75, -2) node[red] {$\bullet$};


  \draw[step = 0.5, dotted] (-1.75, -3) -- (1.75, -3);

  \foreach \x in {-1.75, -1.25, ..., 1.75}
    \draw (\x, -3) node {$\bullet$};


  \foreach \x in {-1.25, -0.75, ..., 1.75} {
    \draw (\x, -3.1) node[anchor = north] {$Z$};
  };

  \draw (-3.5, -3) node {Mixed Dirichlet ($\G\_{D, \dot D}$)};

  \foreach \x in {-1.75, -1.25, ..., 1.25} {
    \draw (\x, -3) node {$\bullet$};
    \draw (\x, -2.9) node[anchor = south] {$X$};
  };

  \draw (-1.75, -3) node[red] {$\bullet$};
  \draw (1.75, -3) node[red] {$\bullet$};
\end{tikzpicture}

\end{center}
\caption{\small Depictions of generating operators for various choices of algebras for spin chains discussed in the text. Algebras generated by these sets are all supported on the entire lattice, i.e.~there are no sites on which all operators act trivially. Red indicates locations of central generators.}
\label{fig scalar algebras}
\end{figure}

It is straightforward to find algebras that correspond to other combinations of open, Neumann, and Dirichlet conditions on edges of the system. For instance, Dirichlet conditions on both edges correspond to the algebra generated by
\bel{
  \G\_{D, D} = \{X_1, \ldots, X_N, Z_2, \ldots, Z_{N - 1}\}.
}
It is also possible to choose algebras where different fields are subject to boundary conditions at the edges. Consider the generating set
\bel{\label{G D Ddot}
  \G\_{D, \dot D} = \{X_1, \ldots, X_{N - 1}, Z_2, \ldots, Z_N\}.
}
The corresponding algebra has a center generated by $X_1$ and $Z_N$, so the density matrices in it correspond to states with definite values of $\phi_1$ and $\dot \phi_N$. This notation is inspired by the fact that $Z_N$ is the momentum operator conjugate to the position $X_N$, so its eigenvalues, schematically, correspond to time derivatives of the field $\phi_N$ at the edge.

The same considerations apply in higher dimensions. Different pairing up of generators on the boundary can be performed to obtain algebras that correspond to fixed derivatives of fields parallel or perpendicular to the boundary. (Links are perpendicular to $\del \Mbb$ if one of their vertices is on $\del \Mbb$ and the other one is not.) It is also straightforward to generalize to bosonic systems with more complicated target spaces: all the operators $Z_v$ and $X_v$ can be replaced by $\Phi_v$ and $\Pi_v$ from eq.\ \eqref{def Phi Pi} or by their inverses, as appropriate.

\subsection{Fermions}

The examples given in this section will focus on fermionic systems in one spatial dimension. As with spin systems, our conclusions can be generalized to higher dimensions, though there will be some additional subtleties we will discuss later. Consider any theory of complex fermions in $d = 1$, defined on $N$ sites labeled $v = 1,\ldots, N$. The maximal algebra is generated by $\{\psi_v, \psi\+_v\}_{v = 1,\ldots, N}$, but unlike the spin operators $X_v$ and $Z_v$, the fermionic basis operators neither form a group nor are they all traceless --- and, most importantly, the generators on different sites all anticommute. This system can, however, be dualized to a bosonic spin system via the \emph{nonlocal} Jordan-Wigner map. By thus bosonizing the fermions, the machinery from the last section can be imported wholesale.

This procedure might seem cumbersome, but matters are greatly simplified by the requirement that the Hamiltonian always be bosonic, i.e.~that all fermionic systems have a fermion parity symmetry, generated by the product of operators $2\psi_v\+\psi_v - 1$ over all sites. If our attention is restricted to states of definite fermion parity and not to their superpositions, then only operators made out of an even number of fermionic generators $\psi_v$ and $\psi\+_v$ will have nonzero expectations. In other words, the algebra of operators for fermionic systems with conserved fermion parity can be taken to be the non-maximal one generated by fermion bilinears $\psi_v\+\psi_v$, $\psi_v \psi_{v + 1}$, and so on. These operators map \emph{locally} under the Jordan-Wigner transformation, even in higher dimensions \cite{Chen:2017fvr, Chen:2018nog}, and the dual algebra (acting on the bosonic Hilbert space of Ising spins) is generated by
\bel{
  \G^f = \{X_1 X_2, X_2 X_3, \ldots, X_{N - 1} X_N, Z_1, \ldots, Z_N\}.
}
We will refer to this algebra as the \emph{maximal fermionic algebra} on $N$ sites, and we will assume that we only work with states of definite fermion parity. The center of this algebra is generated by the fermion parity operator, which is $\prod_{v =1}^N Z_v$ in the bosonic language.

\begin{figure}[tb!]
\begin{center}

\begin{tikzpicture}[scale = 2]

  \draw[step = 0.5, dotted] (-1.75, 0) -- (1.75, 0);

  \foreach \x in {-1.75, -1.25, ..., 1.75}
    \draw (\x, 0) node {$\bullet$};


  \foreach \x in {-1.5, -1, ..., 1.5} {
    \draw (\x, 0.1) node[anchor = south] {$XX$};
  };

  \draw (-3.5, 0) node {Maximal fermionic ($\G^f$)};

  \foreach \x in {-1.75, -1.25, ..., 1.75} {
    \draw (\x, 0) node {$\bullet$};
    \draw (\x, -0.1) node[anchor = north] {$Z$};
  };


  \draw[step = 0.5, dotted] (-1.75, -1) -- (1.75, -1);

  \foreach \x in {-1.75, -1.25, ..., 1.75}
    \draw (\x, -1) node {$\bullet$};


  \foreach \x in {-1.75, -1.25, ..., 1.75} {
    \draw (\x, -1.1) node[anchor = north] {$Z$};
  };

  \foreach \x in {-1, -0.5, ..., 1.5} {
    \draw (\x, -0.9) node[anchor = south] {$XX$};
  };

  \draw (-3.5, -1) node {Dirichlet-open ($\G^f\_{D, O}$)};

  \foreach \x in {-1.75, -1.25, ..., 1.75} {
    \draw (\x, -1) node {$\bullet$};
  };

  \draw (-1.75, -1) node[red] {$\bullet$};


  \draw[step = 0.5, dotted] (-1.25, -2) -- (1.75, -2);

  \foreach \x in {-1.75, -1.25, ..., 1.75}
    \draw (\x, -2) node {$\bullet$};


  \foreach \x in {-1.5, -1, ..., 1.5} {
    \draw (\x, -1.9) node[anchor = south] {$XX$};
  };

  \draw (-3.5, -2) node {Neumann-open ($\G^f\_{N, O}$)};

  \foreach \x in {-0.75, -0.25, ..., 1.75} {
    \draw (\x, -2) node {$\bullet$};
    \draw (\x, -2.1) node[anchor = north] {$Z$};
  };

  \draw (-1.5, -2.1) node[anchor = north] {$ZZ$};

  \draw[dotted, red] (-1.25, -2) -- (-1.75, -2);
  \draw (-1.25, -2) node[red] {$\bullet$};
  \draw (-1.75, -2) node[red] {$\bullet$};


  \draw[step = 0.5, dotted] (-1.75, -3) -- (1.75, -3);

  \foreach \x in {-1.75, -1.25, ..., 1.75}
    \draw (\x, -3) node {$\bullet$};

  \foreach \x in {-1.5, -1, ..., 1.5} {
    \draw (\x, -2.9) node[anchor = south] {$XX$};
  };


  \foreach \x in {-1.25, -0.75, ..., 1.75} {
    \draw (\x, -3.1) node[anchor = north] {$Z$};
  };

  \draw (-3.5, -3) node {Majorana-open ($\G^f\_{M, O}$)};

  \foreach \x in {-1.75, -1.25, ..., 1.25} {
    \draw (\x, -3) node {$\bullet$};
  };

  \draw (-1.75, -3) node[red] {$\bullet$};
\end{tikzpicture}

\end{center}
\caption{\small Generators of fermionic algebras discussed in the text, presented using their bosonized equivalents. Each algebra commutes with the product of all $Z$'s, i.e.~with the total fermion parity $(-1)^F$. Locations of all \emph{additional} central generators are red. In the case of Majorana boundary conditions, the removal of $Z_1$ is taken to mean that $(-1)^F$ is still a central generator, but that an overall constraint is imposed on superselection sector weights such that $(-1)^F = 0$ hold as an operator equation.}
\label{fig fermion algebras}
\end{figure}
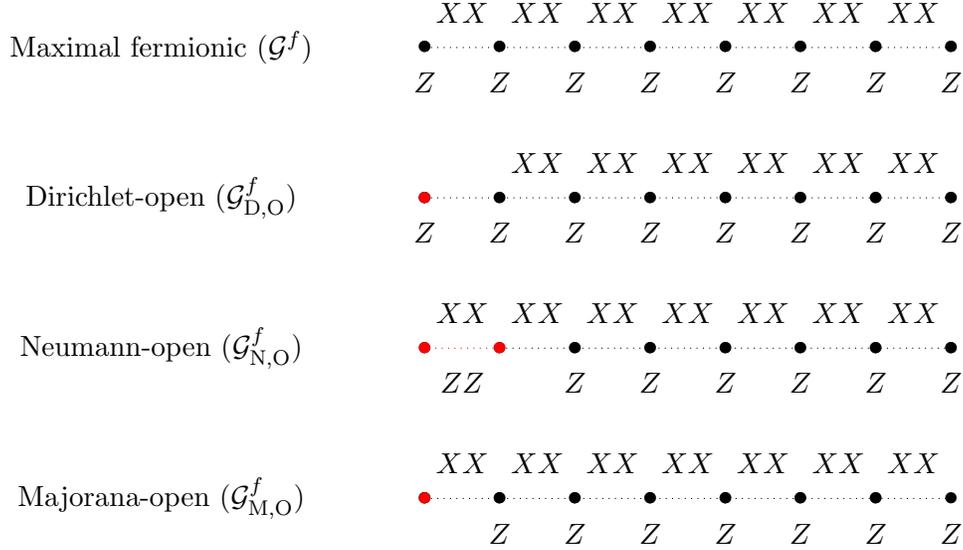

What edge modes or boundary conditions can be imposed here? A natural first step is the removal of the operator $X_1X_2$ from $\G^f$, resulting in the generating set
\bel{
  \G^f\_{D,O} = \{X_2 X_3, \ldots, X_{N - 1} X_N, Z_1, \ldots, Z_N\}.
}
This adds another central generator, $Z_1$, to the existing one (the fermion parity). The extra su\-per\-se\-lec\-tion sectors that ensue are labeled by the number of fermions on the edge. This is the fermionic dual to Dirichlet boundary conditions.

The other natural alternative is
\bel{
  \G^f\_{N,O} = \{X_1 X_2, \ldots, X_{N - 1} X_N, Z_1 Z_2, Z_3, \ldots, Z_N\}.
}
The additional central generator here is $X_1 X_2$, just like in the Neumann boundary condition for spins. In the fermionic language this central generator is rather nontrivial: it corresponds to $(\psi_1\+ - \psi_1)(\psi_2\+ + \psi_2)$.  However, in terms of Majorana fermions $\chi_v$ and $\chi'_v$, defined via $\psi_v = \chi_v + \i \chi_v'$, this central generator is simply the Majorana hopping operator $\i \chi_1'\chi_2$. (In this Majorana basis, the central generator corresponding to Dirichlet conditions is $\i\chi_1'\chi_1$.)

It is interesting to examine the generating set obtained by removing $Z_1$ from $\G^f$,
\bel{
  \G\_{M, O}^f = \{X_1X_2, \ldots, X_{N - 1}X_N, Z_2, Z_3, \ldots, Z_N\}.
}
This is a natural object in terms of Majorana generators $\chi_v$ and $\chi_v'$ introduced in the preceding paragraph, as the corresponding algebra is generated by all Majorana bilinears built \emph{without} $\chi_1$.\footnote{It is also possible to remove bilinears involving $\chi_1'$ from the generating set, which corresponds to removing both $X_1 X_2$ and $Z_1$ from $\G^f$. However, the resulting algebra is not supported on all sites --- there are no nontrivial operators acting on site 1 --- so we do not consider this case here.} This algebra has no center. Moreover, it does not contain the fermion parity operator $(-1)^F$. Thus, it only describes states with $(-1)^F = 0$ and so it cannot contain the density matrix of any state with definite fermion parity; it can only contain equal statistical mixes of states with $(-1)^F = \1$ and $(-1)^F = -\1$.  We can say that the algebra generated by $\G\_{M, O}^f$ corresponds to ``Majorana boundary conditions.'' We will return to such kinds of boundary conditions when discussing algebras on subregions.

The notion of Dirichlet and Neumann subalgebras of the maximal fermionic algebra can be extended to higher dimensions. The procedure is the same in spirit as the one outlined above: the algebra of fermion bilinears is dualized to a (non-maximal) subalgebra acting on a $\Z_2$ bosonic system, and appropriate subalgebras with additional central generators at the edges are then considered. The complication is that duals of fermionic systems in higher dimensions are gauge theories with non-standard Gauss laws \cite{Chen:2017fvr, Chen:2018nog}. Nevertheless, the fermion bilinears all dualize into local operators, and the procedure is conceptually straightforward and merely technically involved.

\subsection{Abelian gauge theories} \label{subsec gauge theories}

Let us now consider a $\Z_2$ gauge theory in $d = 2$. (Generalizations to $\Z_K$ and $U(1)$ theories are straightforward.) To recap, the full Hilbert space is a product of $K = 2$ spaces on each link $\ell$ of the lattice $\Mbb$. The algebra $\A$ of gauge-invariant operators, described in section \ref{sec formal}, is generated by electric fields $X_\ell$ on links, magnetic fields $W_f$ on faces, and Wilson loops $W_c$ along homologically distinct noncontractible 1-cycles. Pure density matrices in $\A$ must describe eigenstates of Gauss operators $G_v = \prod_{\ell\supset v} X_\ell$ on vertices, as these operators generate the center of $\A$. Gauge-invariant states are those with $G_v = \1$ for all $v$.

What are natural ways to reduce this algebra by removing operators at the edges of the system? A subalgebra of $\A$ without some of the Gauss operators will not contain density matrices of any pure gauge-invariant states; it will necessarily only contain mixtures of states with $G_v = \1$ and $G_v = -\1$ for those Gauss operators $G_v$ absent from the subalgebra. This is analogous to what happened with fermionic subalgebras that did not contain the fermion parity operator. If our goal is to formulate gauge theory subalgebras that correspond to boundary conditions on \emph{pure} states, the subalgebras must then contain all Gauss operators. Appropriate subalgebras will thus be obtained either by removing magnetic generators or by removing electric generators while keeping at least those products of theirs that form Gauss operators.

\begin{figure}
\begin{center}

\begin{tikzpicture}[scale = 2]

  \draw[step = 0.5, thick] (-1, -1) grid (1, 1);
  \draw[blue, thick] (-1, -1) rectangle (1, 1);
  \filldraw[fill=green!20, draw=green!50!black, thick]  (0.5, 0) rectangle +(0.5, 0.5);
  \draw[very thick, red] (1, 0) -- (1, 0.5);

  \filldraw[fill=white, white] (-0.65, -0.65) rectangle +(0.3, 0.3);
  \draw (-0.5, -0.5) node {$\Mbb$};
  \draw (-1, 0.5) node[anchor = east, blue] {$\del \Mbb$};

  \draw (0.75, 0.25) node[green!60!black] {$W_{f_1}$};
  \draw (1, 0.25) node[red, anchor = west] {$X_{\ell_1}$};

\end{tikzpicture}

\end{center}
\caption{\small A $\Z_2$ gauge theory on a lattice $\Mbb$ with boundary $\del \Mbb$ (drawn in blue). There is a central generator $G_v$ at every vertex of this lattice, including on edge sites $v \in \del \Mbb$. Boundary plaquettes are those faces that contain boundary links; for instance $f_1$ contains a boundary link $\ell_1 \in \del \Mbb$. When $W_{f_1}$ is removed from the algebra, $X_{\ell_1}$ becomes another central generator. It corresponds to the electric field operator parallel to the boundary.}
\label{fig gauge-inv algebras}
\end{figure}
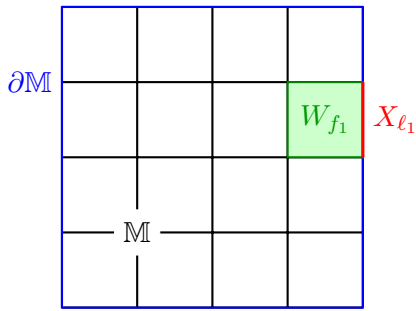

Consider first the generating set $\G_{\trm{E}(\ell_1)}$ obtained by removing the magnetic field operator $W_{f_1}$ from a plaquette $f_1$ that contains the link $\ell_1\in \del\Mbb$, see fig.~\ref{fig gauge-inv algebras}. The removal of $W_{f_1}$ causes the operator $X_{\ell_1}$ to become a central generator. Thus pure states whose density matrices are in $\A_{\trm{E}(\ell_1)}$ have a definite electric field on the boundary link $\ell_1$. We can similarly remove magnetic field operators $W_{f_2}$, $W_{f_3}$,\ldots from other plaquettes that contain boundary links $\ell_2$, $\ell_3$, and so on. If all the magnetic operators from boundary faces are removed, we will say that the remaining generators generate the \emph{electric algebra}, and will denote it by $\A\_E$. This algebra contains density matrices of all states with definite electric fields at the boundary. In the more common continuum notation, these are states with definite values of the electric field $E_\parallel$ parallel to the boundary. Note that because of the Gauss law at each boundary site, knowing $E_\parallel$ as a function of the position along the boundary is the same as knowing $E_\perp$, the electric field flowing into the boundary at each boundary site.

The analysis of what happens when certain electric field operators are removed from $\A$ is a bit more involved. The simplest way to obtain a nontrivial subalgebra that contains all the boundary Gauss operators is to consider the generating set $\G\_M$ that contains none of the individual $X_\ell$ operators on the boundary of the lattice, but that does contain all the Gauss operators $G_v$ for edge sites $v$. Pure states whose density matrices are in this algebra $\A\_M$ are eigenstates of the Wilson loop along the boundary of the region; in continuum notation, this corresponds to a fixed value of $\oint A$ along the boundary. These states hence have definite values of total magnetic flux through the entire spatial surface. We will call the corresponding $\A\_M$ the \emph{magnetic algebra}.

Other admissible subalgebras are obtained by removing more electric operators from plaquettes $f_i$ on the boundary in such a manner that every boundary plaquette has only products of an even number of electric operators acting on its links. This way individual magnetic fields $W_{f_i}$ on the boundary can become generators of the center. The pure density matrices in such algebras $\A_{\trm{M}(f_i)}$ correspond to states with definite magnetic fields at the boundary.

We introduced Neumann and Dirichlet boundary conditions for 1d fermions as Jordan-Wigner duals of those boundary conditions for bosons. Given that 2d Abelian gauge theories can be dualized to bosonic theories via the Kramers-Wannier transform, it may appear natural to compare the electric and magnetic conditions of $\Z_2$ theories to Dirichlet and Neumann conditions in dual spin systems. However, this is not straightforward. A gauge theory on a lattice with boundary is dual to a spin system only if it contains no operators on the boundary --- Gauss operators included \cite{Radicevic:2016tlt, Moitra:201x}. A duality between boundary conditions can indeed be established, but the paradigm developed in this section would need to be altered to account for gauge theories without boundary operators. We will not explore this issue here.

Finally, we note that in relativistic gauge theories there exists another terminology for boundary terms. Any boundary condition that fixes the field strength $F_{\mu\nu}$ with $\mu$ and $\nu$ both parallel to the boundary is a Dirichlet condition, while fixing $F_{\mu\nu}$ with $\mu$ or $\nu$ perpendicular to the boundary is a Neumann condition. This means that our magnetic conditions can be either Neumann or Dirichlet in $d \geq 3$; in $d = 2$ magnetic conditions can only be Neumann, and in $d = 1$ they are not defined. Our electric boundary conditions can only be Dirichlet in $d \geq 2$ (if we are fixing $X_\ell$ for $\ell \in \del \Mbb$) but can always be Neumann (if we are fixing $X_\ell$ for $\ell \perp \del \Mbb$).

\vfill

\noindent \begin{minipage}{\linewidth} 
Table 1: {\small A summary of correspondences between algebras and boundary conditions in scalar, fermion, and (Abelian) gauge theories. Most common boundary conditions are listed.}
\begin{footnotesize}
\begin{center}
\begin{tabular}{|c|ccc|} \cline{2-4}
   \multicolumn{1}{c|}{} & Generators of the algebra & Generators of the center & \multicolumn{1}{c|}{Boundary conditions} \\ \hline
  Spin chains
  & $X_1,\ldots, X_N, Z_1,\ldots, Z_N$ & None & Open \\ \cline{2-4}
    & $X_1, \ldots, X_N, Z_2,\ldots, Z_{N - 1}$
    & $X_1, X_N$
    & Dirichlet \\ \cline{2-4}
  & \makecell{$X_1, \ldots, X_N$,\\ $Z_1Z_2, Z_3, \ldots, Z_{N- 2}, Z_{N - 1} Z_N$}
    & $X_1X_2, X_{N - 1}X_N$
    & Neumann \\ \hline \hline
  Fermion chains
  & \makecell{$\chi_1' \chi_1, \chi_2'\chi_2, \ldots, \chi_N' \chi_N$, \\ $\chi_1'\chi_2, \chi_2' \chi_3, \ldots, \chi_{N - 1}'\chi_N$}
    & $(-1)^F \equiv (\i\chi_1'\chi_1) \cdots (\i\chi_N' \chi_N)$
    & Open \\ \cline{2-4}
  & \makecell{$\chi_1' \chi_1, \ldots, \chi_N' \chi_N$, \\ $\chi_2'\chi_3, \ldots, \chi_{N - 2}'\chi_{N - 1}$}
    & $(-1)^F, \chi_1'\chi_1, \chi_N' \chi_N$
    & Dirichlet \\ \cline{2-4}
  & \makecell{$\chi_1'\chi_1 \chi_2'\chi_2, \chi_{N - 1}' \chi_{N - 1} \chi_N' \chi_N$,\\ $\chi_3'\chi_3, \ldots, \chi_{N - 2}'\chi_{N - 2}$, \\ $\chi_1' \chi_2, \chi_2'\chi_3, \ldots, \chi_{N - 1}'\chi_N$}
    & $(-1)^F, \chi_1'\chi_2, \chi_{N - 1}' \chi_N$
    & Neumann \\
  \hline \hline
  $\Z_2$ gauge theory
  & $X_\ell, W_f \equiv \prod_{\ell\subset f} Z_\ell$ for all $\ell$, $f$
    & $G_v \equiv \prod_{\ell\supset v} X_\ell$ for all $v$
    & Open\\ \cline{2-4}
  (any dimension)
  & All $X_\ell, W_f$ except $W_f$, $f \in \del \Mbb$
    & All $G_v$, and $X_\ell$ for $\ell \in \del \Mbb$
    & Electric\\ \cline{2-4}
  & All $X_\ell, W_f$ except $X_\ell$, $\ell \in \del \Mbb$
    & All $G_v$, and $W\_{\del \Mbb} \equiv \prod_{\ell \in \del\Mbb} Z_\ell$
    & Magnetic\\ \hline \hline
  Compact scalar
  & $\Phi_v \equiv \e^{\i \phi_v}, \Pi_v$ for all vertices $v$
    & None
    & Open\\ \cline{2-4}
    (any dimension)
  & All $\Phi_v, \Pi_v$ except $\Pi_v$ for $v \in \del \Mbb$
    & All $\Phi_v$ for $v \in \del \Mbb$
    & Dirichlet\\ \cline{2-4}
  & \makecell{All $\Phi_v$, all $\Pi_v$ in the bulk,\\ only $\Pi_v \Pi_u$ for links $(v,u) \perp \del \Mbb$}
    & All $\Phi_v^{-1} \Phi_u$ for $(v,u) \perp \del \Mbb$
    & Neumann \\ \hline \hline
  $U(1)$ gauge theory
  & $X_\ell \equiv \e^{\pder{}{A_\ell}}, W_f \equiv \e^{\i \oint_f A}$ for all $\ell$, $f$
    & $G_v \equiv \e^{\oint_v * \pder{}A}$ for all $v$
    & Open\\ \cline{2-4}
  (any dimension)
  & All $X_\ell, W_f$ except $W_f$, $f \in \del \Mbb$
    & All $G_v$, and $X_\ell$ for $\ell \in \del \Mbb$
    & Electric\\ \cline{2-4}
  & All $X_\ell, W_f$ except $X_\ell$, $\ell \in \del \Mbb$
    & All $G_v$, and $W_f$ for $f \in \del \Mbb$
    & Magnetic\\ \hline
\end{tabular}
\end{center}
\end{footnotesize}
\end{minipage}

\subsection{Nonabelian gauge theories} \label{subsec nonabelian gauge theories}

The direct approach indicated in section \ref{sec formal} is not very convenient for nonabelian gauge theories, so we adjust it as follows \cite{Ghosh:2015iwa, Radicevic:2015sza}. We start from a theory in which on each link $\ell$ we have a Hilbert space $\H_\ell$ spanned by $\{\qvec U_\ell\}$ for all $U$ in a given gauge group $G$. Now, on each link we define an infinite series of finite-dimensional Hilbert spaces, $\H^R_\ell$, associated to irreducible representations $R$ of $G$. A basis of $\H^R_\ell$ is
\bel{\label{def R alpha beta}
  \qvec{R_{\alpha\beta}}_\ell \propto \int_G \d U R_{\alpha\beta}(U) \qvec U_\ell, \quad \alpha, \beta = 1, \ldots, \dim R.
}
Here we omit normalization factors and take $\d U$ to be the Haar measure on $G$. The dimension of $\H^R_\ell$ is $(\dim R)^2$. The vectors $\{\qvec{R_{\alpha\beta}}_\ell\}$ for all $R$, $\alpha$, $\beta$ span $\H_\ell$.\footnote{A finite approximation to a nonabelian gauge theory can be obtained by keeping only representations below a certain dimension, but then the algebra of operators may only approximately close under multiplication.}

States $\qvec{R_{\alpha\beta}}_\ell$ have definite electric flux in representation $R$, but their products over all $\ell$ are generally not gauge-invariant. Gauss operators that implement gauge transformations by $\Lambda \in G$ are
\bel{
  G_v^\Lambda = \prod_u L^\Lambda_{vu},
}
where the product goes over all neighboring vertices of $v$, and $L^\Lambda_{vu} \qvec U_\ell$ equals $\qvec{\Lambda U}_\ell$ or $\qvec{U \Lambda^{-1}}_\ell$, depending on whether the link $\ell$ is oriented from $v$ to $u$ or vice versa.\footnote{In terms of states $\qvec{R_{\alpha\beta}}$, these two types of operators act on $\alpha$ and $\beta$ by the matrices $R(\Lambda^{-1})$ and $R^T(\Lambda)$, respectively. For example, $L^\Lambda_\ell \qvec{R_{\alpha\beta}}_\ell \propto \int_G \d U \, R_{\alpha\alpha'}(\Lambda^{-1}) R_{\alpha' \beta}(U) \qvec U_\ell \equiv \qvec{R_{\alpha\alpha'}^{-1}(\Lambda) R_{\alpha'\beta}}_\ell$. } The operators $L^\Lambda_{vu}$ are nonabelian generalizations of operators built out of electric flux generators $X_\ell$ from the previous subsection, but unlike them $L^\Lambda_{vu}$ are not gauge-invariant. It is common to view $G$ as a Lie group and to employ its generators $T^a$ to write
\bel{\label{def J}
  L_{vu}^\Lambda = \e^{\i \theta^a J^a_{vu}} \quad\trm{for}\quad \Lambda = \e^{\i \theta^a T^a},
}
with the understanding that the electric flux generators $J^a_{vu}$ act either from left or from right and have sign $+$ or $-$, depending on the orientation of the link \cite{Ghosh:2015iwa}.

Gauge-invariant states are products of states $\qvec{R_{\alpha\beta}}_\ell$ in the same representation along a cycle, with indices $\alpha, \beta, \ldots$ contracted between neighboring links in this cycle. Gauge-invariant operators are Wilson loops $W^R_c$ in representations $R$ on cycles $c$, and Casimirs built out of electric flux generators $J_{uv}$ defined in eq.~\eqref{def J}. The analysis of boundary conditions in non-maximal algebras now proceeds analogously to the Abelian case: removing e.g.~all $W^R_f$ from a face $f$ near $\del \bb V$ turns all the Casimirs on the boundary link from $f$ into central generators. Conversely, removing all Casimirs and local Wilson loops  from $\del \bb V$ makes all the $W_{\del \bb V}^R$'s become central.

\section{Subalgebras and boundary conditions} \label{sec subalgebras}

The analysis so far has dealt with algebras that are supported on the whole lattice $\Mbb$, and a particular emphasis was placed on understanding which pure density matrices belong to which algebras. In this section we will deal with algebras $\A_{\bb V}$ of operators supported on a \emph{subregion} $\bb V \subset \Mbb$. We assume that we are given a density matrix $\rho$ in the full algebra $\A$ --- or, equivalently, a list of expectation values $\avg \O$ of all basis elements $\O \in \A$. Each pair $(\rho, \A_{\bb V})$ uniquely induces a reduced density matrix via an analog of eq.~\eqref{def rho},
\bel{
  \rho(\A_{\bb V}) \equiv \frac1D \sum_{a'} \avg{\O^{-1}_{a'}} \O_{a'}, \quad \trm{where}\quad \A_{\bb V} = \trm{span}\{\O_{a'}\}.
}
We seek to understand what kinds of different $\rho(\A_{\bb V})$'s can there be, given a state $\rho$ and a region $\bb V$.

We will focus on algebras $\A_{\bb V}$ that differ by operators at the edges of $\bb V$, and the resulting set of different possible $\rho_{\bb V}$ will correspond to different boundary conditions that can be put at the entangling cut of the system. This is analogous to section \ref{sec algebras}. Our analysis applies more broadly, and the same techniques can be applied to algebras $\A_{\bb V}$ lacking generators in the interior of $\bb V$.


Any reduced density matrix in any subalgebra $\A_{\bb V}$ has the form
\bel{
  \rho(\A_{\bb V}) = \rho_{\bb V} \otimes \left(\frac1{D_{\bar{\bb V}}} \1_{\bar{\bb V}}\right),
}
where $\rho_{\bb V}$ is a properly normalized density matrix of a state in $\H_{\bb V}$ (the product of all target Hilbert spaces $\H_0$ associated to elements of $\bb V$ that carry degrees of freedom), and $\1_{\bar{\bb V}}$ acts only on the complement $\H_{\bar{\bb V}}$ of dimension $D_{\bar{\bb V}}$. This means that the algebra $\A_{\bb V}$ only contains density matrices of states that are uniform mixtures of all possible field configurations outside $\bb V$. From now on we focus only on $\rho_{\bb V}$, for which there are three possibilities:
\begin{enumerate}
  \item If $\A_{\bb V}$ is the maximal algebra $\C^{D_{\bb V} \times D_{\bb V}}$ on $\H_{\bb V}$, there will be open boundary conditions on the edges of the interval, and all density matrices $\rho_{\bb V}$ will be allowed.
  \item If $\A_{\bb V}$ is not maximal but has a center generated by edge operators, $\rho_{\bb V}$ will split into superselection sectors that contain states obeying a specific kind of boundary condition. Examples include Dirichlet and Neumann conditions from the previous section. Relative weights of different sectors depend on the original state and may take arbitrary values.
  \item If $\A_{\bb V}$ is not maximal and has no center generated by edge operators, we can add a central generator and require that its expectation value always be zero. The reduced density matrix will then split into blocks labeled by eigenvalues of this central generator, but there will be a constraint on the relative weights of superselection sectors that will force  $\rho_{\bb V}$  to be impure. An example is the Majorana boundary condition from the previous section. In this case the field at the boundary is \emph{random}, i.e.~it is an equal statistical mixture between states with and without a fermion at the edge.
\end{enumerate}

Boundary conditions may effectively depend on $\rho$ in the following sense. If $\A_{\bb V}$ is the maximal algebra on $\bb V$, and if any operator that fails to commute with a given $\O\in \A_{\bb V}$ has zero expectation, then $\rho_{\bb V}$ will be block-diagonal with sectors labeled by eigenvalues of $\O$. For instance, if $\Pi$ is a momentum operator conjugate to a position operator $\Phi$, open boundary conditions are the same as Dirichlet conditions in a state where all expectations involving $\Pi$ vanish (i.e.~if $\Pi = 0$). As another example, if $\A_{\bb V}$ is the Dirichlet algebra of a spin system with $Z_v$ a central generator, and if the state has $Z_v = 0$ for some $v \in \del {\bb V}$, the Dirichlet boundary conditions will be the same as random (Majorana) ones: only a uniform mix of $Z_v = \1$ and $Z_v = -\1$ states will appear in $\rho_{\bb V}$. When specifying a set of boundary conditions, we will always specify them based on the algebra and not on the particular state in question.

Let us now apply these ideas to our usual test subject, the $\Z_2$ gauge theory. Gauge invariance has historically caused a lot of confusion when it came in contact with the study of entanglement. By the following analysis we wish to very forcefully point out that even when working with the full Hilbert space the results will be gauge-invariant, as long as only gauge-invariant algebras are studied. These points have been made in \cite{Radicevic:2016tlt}.

Consider an algebra $\A_{\bb V}$ of gauge-invariant operators supported on a set $\bb V$, which in this case is a collection of links. No operator in $\A_{\bb V}$ has support outside of $\bb V$ --- in particular, no Gauss operator $G_v$ for $v \in \del\bb V$ is in $\A_{\bb V}$.\footnote{We assume that $\bb V$ is a subregion of $\Mbb$ that is far away from the boundary $\del \Mbb$ of our spatial lattice, if any exists.} The center of $\A_{\bb V}$ is generated by Gauss operators in the interior of $\bb V$ and by boundary electric fields
\bel{\label{def E abel}
  E_v \equiv \prod_{\substack{\ell\supset v\\ \ell \in \bb V}}X_\ell.
}
If the original state $\rho$ is gauge-invariant, then all Gauss operators will have $G_v = \1$, and the appropriate reduced density matrix $\rho_{\bb V}$ will contain only states that satisfy the Gauss law on all the interior points, i.e.~it will be of the form
\bel{\label{rho gauge}
  \rho_{\bb V}  = \~\rho_{\bb V} \prod_{v \in \trm{Int}(\bb V)} \frac{\1 + G_v}2.
}

The remaining central generators of $\A_{\bb V}$, $E_v$, ensure that the generic density matrix $\rho_{\bb V}$  describes a mixture of states with definite electric fluxes through $\del \bb V$. This \emph{almost} corresponds to electric boundary conditions on $\del\bb M$ as described in the previous section. Recall that our original definition of electric boundary conditions involved specifying all electric fields on links in $\del\Mbb$. In that situation, all magnetic field generators were removed from plaquettes containing boundary links, and $W_{\del \bb M}$, the Wilson loop along the boundary of $\bb M$, was not in the algebra of operators dubbed $\A\_E$. Here, however, all magnetic operators on the boundary are retained, and in particular $W_{\del \bb V} \in \A_{\bb V}$. Thus, with electric boundary conditions on physical boundaries of the system, density matrices split into $2^{|\del \bb V|}$ blocks; with electric boundary conditions on the entanglement edge, they split into $2^{|\del \bb V| - b_0}$ blocks, where $b_0$ is the number of disconnected components of $\bb V$. We will still refer to these boundary conditions as electric, but this caveat must be kept in mind, as the $b_0$ ``correction'' gives rise to universal quantities such as the topological entanglement entropy.\footnote{Non-maximal gauge-invariant algebras, obtained by removing gauge-invariant operators from the vicinity of $\del \bb V$ in a manner described in section \ref{subsec gauge theories}, can also be viewed as having random boundary conditions for some of the central generators $E_v$. We will not use this point of view in this paper, but it may be a useful fact when comparing entropies of different algebras in a fixed quantum state.}

In nonabelian theories, the center of the gauge-invariant algebra is also given by the Gauss operators in the interior and by a set of operators on the boundary of $\bb V$. These boundary operators are defined as Casimirs built out of products over all interior vertices $u$ neighboring $v \in \del \bb V$,
\bel{\label{def E Lambda}
  E_v^\Lambda \equiv \prod_{u \in \bb V} L_{vu}^\Lambda.
}
How do these operators act in the basis $\qvec{R_{\alpha\beta}}_\ell$ from eq.~\eqref{def R alpha beta}? Within each $\H^R_\ell$ such that $v \in \ell \in \bb V$, we have
\bel{
  L_{vu}^\Lambda: \qvec{R_{\alpha\beta}}_\ell \mapsto \qvec{R_{\alpha\alpha'}^{-1}(\Lambda) R_{\alpha'\beta}}_\ell \quad \trm{or} \quad \qvec{R_{\alpha\beta'} R_{\beta'\beta}(\Lambda)}_\ell,
}
depending on the orientation of $\ell$ (recall that we are summing over repeated indices, and when the argument of $R$ is dropped it is understood to be the integration variable $U$ appearing in \eqref{def R alpha beta}).  The product of these operators, $E^\Lambda_v$, has an analogous action on the \emph{total} flux in representation $R$ that enters the vertex $v$ \cite{Ghosh:2015iwa}. Thus, every central operator $E_v^\Lambda$ in a nonabelian theory has an identity component when represented on the space $\bigotimes_\ell \H^R_\ell$, and hence the reduced density matrix will contain this product of identities, too. (We will give them a physical interpretation in section \ref{s3}.) Beyond this, the analysis of nonabelian theories is analogous to the Abelian case: the existence of boundary central generators causes $\rho_{\bb V}$ to split into superselection sectors labeled by the tuple $R$, which are direct generalizations of Abelian electric fluxes labeled by $k$.

\section{Entropies of reduced density matrices} \label{sec entropies}

The algebraic point of view developed in the previous sections allows us to start from any theory with a known lattice regularization and define the Hilbert space in the natural representation of any subalgebra of operators $\A_{\bb V}$. Roughly speaking, this space is the direct product $\H_{\bb V} = \bigotimes_{i \in \bb V} \H_i$ over all degree-of-freedom-carrying simplices $i$ (vertices, links, faces, etc) on which at least one operator in $\A_{\bb V}$ acts nontrivially. More precisely, a non-maximal algebra $\A_{\bb V}$ restricts the possible quantum superpositions between the states in $\H_{\bb V}$, so the resulting restricted space will be denoted $\H^*_{\bb V}$. This restricted space $\H^*_{\bb V}$ has the same dimension as $\H_{\bb V}$. The only difference between them is that $\H^*_{\bb V}$ does not allow superpositions of states in different superselection sectors. If $k$ labels these sectors, we can write $\H^*_{\bb V} = \bigoplus_k \H^{(k)}_{\bb V}$, where each $\H^{(k)}_{\bb V}$ is a smaller but more ``traditional'' kind of Hilbert space where all superpositions are allowed.

The natural entropy to be associated with $\rho_{\bb V}$ is the von Neumann entropy,
\bel{\label{def S vN}
  S\_{\bb V} = - \Tr\left(\rho_{\bb V} \log \rho_{\bb V}\right) = -\pder{}n \bigg|_{n = 1} \Tr \rho_{\bb V}^n.
}
The trace is taken over $\H_{\bb V}$ or, equivalently, over $\H_{\bb V}^*$.\footnote{It is possible to take other definitions of the trace. For instance, it may be over the smallest faithful representation of $\A_{\bb V}$, or it may be over all of $\H$. Different choices will lead to entropies that differ by state-independent constants. It is important to only compare entropies defined using the same trace.} The latter formula is the basis of the replica trick, to be explored in the next section. We will call this quantity the \emph{full entanglement entropy}.

We again stress that in a lattice gauge theory, we take $\H_{\bb V}$ to be the set of all states on links in $\bb V$ --- gauge-invariant or otherwise. The projectors in \eqref{rho gauge} make sure that only states obeying $G_v = \1$ for $v \in \trm{Int}(\bb V)$ contribute to the trace. More interestingly, $\rho_{\bb V}$ does not contain projectors that enforce $G_v = \1$ for $v \in \del \bb V$. This is an important point whose ramifications we study below eq.~\eqref{S full} and in section \ref{s3}.

If $\rho_{\bb V}$ is block diagonal in some basis, we can first compute $\Tr\left(\bigoplus_k \rho_{\bb V}^{(k)}\right)^n = \sum_k \Tr \left(\rho_{\bb V}^{(k)}\right)^n$ and then write
\bel{\label{def S replica}
  S_{\bb V} = -\pder{}n \Bigg |_{n = 1} \left[\sum_k \Tr \left(\rho_{\bb V}^{(k)}\right)^n \right].
}
If there is only one relevant superselection sector, i.e.~if $\A_{\bb V}$ is the maximal algebra on $\H_{\bb V}$ or if all sectors but one have zero density matrices, this formula reduces to \eqref{def S vN}.

In the presence of superselection sectors, the entropy $S_{\bb V}$ is not the only interesting quantity to define.  Let us first assume that we are working with matter or Abelian gauge theories. It is useful to define unit-trace density matrices $\bar \rho$ in each sector:
\bel{\label{def pk}
  \rho_{\bb V}^{(k)} = p_k \, \bar\rho_{\bb V}^{(k)}, \quad p_k \equiv \Tr \rho_{\bb V}^{(k)}.
}
The entropy associated to $\A_{\bb V}$ is now
\bel{\label{entropy p}
  S_{\bb V} = - \sum_k p_k \log p_k + \sum_k p_k S^{(k)}_{\bb V},
}
with
\bel{\label{def SVk}
   S_{\bb V}^{(k)} = - \pder{}n \bigg|_{n = 1} \Tr\left(\bar\rho_{\bb V}^{(k)}\right)^n.
}
The quantity
\bel{
  S_{\bb V}\^{dist} \equiv \sum_k p_k S^{(k)}_{\bb V}
}
is called the \emph{distillable entanglement entropy}.\footnote{The edge terms are not distillable using gauge-invariant operations in the interiors of the entangling region \cite{Ghosh:2015iwa}, i.e.~they are not convertible to entangled Bell pairs in an external reservoir of qubits via the local operations. The obstruction is that gauge-invariant operators in the interiors of the regions cannot change the superselection sector. See \cite{Soni:2015yga} for an explicit distillation protocol in a constrained spin system that models a lattice gauge theory.} For a generic state, it is different from the full entropy $S_{\bb V}$ whenever $\A_{\bb V}$ has central generators.

In case we are working with nonabelian gauge theories, recall from the discussion around eq.~\eqref{def E Lambda} that $\rho_{\bb V}$ has more structure beyond splitting into sectors labeled by the representation $R$ of the gauge group: within each sector, it is a tensor of a $\dim R \times \dim R$ identity matrix and a nontrivial matrix that we may call the proper reduced density matrix. Each representation $R$ thus contributes an entropy of $\log \dim R$ to $S^{(k)}_{\bb V}$. (Here we are using $k$ to denote all labels for superselection sectors; in general, $R$ will be a subset of $k$, meaning that there may be multiple sectors $k$ with the same label $R$.) In fact, it is natural to define $S^{(k)}_{\bb V}$ as the entropy of the proper reduced density matrix in sector $k$, as the  $\log\dim R$ contribution is also not distillable. Hence, in a nonabelian gauge theory, we will decompose the full entanglement entropy as
\bel{\label{entropy p R}
  S_{\bb V} = - \sum_k p_k \log p_k + \sum_k p_k S^{(k)}_{\bb V} + \sum_R p_R \log \dim R,
}
where $p_R$ is the total weight of all sectors $k$ that contain $R$.

With this convention,
\bel{\label{def S ginv}
  S_{\bb V}\^{g\textrm{-}inv} \equiv - \sum_k p_k \log p_k + \sum_k p_k S^{(k)}_{\bb V}
}
will be called the \emph{gauge-invariant entropy} of $\bb V$. This is the entropy that we would get if the trace in \eqref{def S vN} was taken in the representation associated to gauge-invariant degrees of freedom only, instead of using the representation on the full space $\H_{\bb V} = \bigotimes_\ell \H_\ell$. We caution the reader that this name is slightly misleading, as the full entropy \eqref{entropy p R} is gauge-invariant, too.

\section{From subalgebras to path integrals} \label{sec path integrals}

We will now show how the various entropies defined in the previous section can be computed using path integrals.\footnote{\label{foot A0 role} When working with theories with finite target spaces, like spin systems or fermions, we may either consider the time $\tau$ to be discrete, or we may take it to be continuous with the understanding that all integrals are constructed using the coherent state approach \cite{Altland:2010}. Either way, $\tau$ is always Euclidean. For gauge theories, it is most helpful to keep $\tau$ discrete, with fields $A_0$ living on temporal links. Integrating out the $A_0$'s in any Maxwell/Yang-Mills theory enforces local conservation of charge, making the path integral include only gauge-invariant configurations \cite{Kogut:1974ag}.} Let $\left\{\left\qvec{\phi^{(k)}\right}\right\}$ be a basis of $\H^*_{\bb V}$ that diagonalizes all central elements of $\A_{\bb V}$, and let $\{\qvec{\vartheta}\}$ be any basis of $\H_{\bar{\bb V}}$. The eigenvalues of central generators are collected in the tuple $k$. For instance, in a scalar theory with Dirichlet boundary conditions on the entangling edge, $\phi^{(k)}_i$ could be used to label the position eigenstate with field configuration $\phi_i$ for $i \in \trm{Int}(\bb V)$ and with boundary conditions $\phi_v = k_v$ for $v \in \del \bb V$. In a gauge theory, instead of including in $k$ the Gauss operator eigenvalues in $\trm{Int}(\bb V)$ (which are all equal to unity, cf.~\eqref{rho gauge}), we will simply require that any $\qvec{\phi^{(k)}}$ be invariant under all gauge transformations in Int$(\bb V)$. This way in gauge theories the tuple $k$ will contain only eigenvalues of central operators on the edge.

The wave function of the ground state $\qvec\Psi$ of a theory with Euclidean Lagrangian $L$ is given by
\bel{\label{def Psi}
\begin{gathered}
  \Psi_k(\phi, \vartheta)
  \equiv \qprod{\phi^{(k)}, \vartheta \Big}{\Psi} = \int^{(k)} [\d \varphi]\, \e^{-\int_{-\infty}^0 L\, \d \tau },\\
  \varphi_i(\tau = 0^-) = \phi^{(k)}_i \ \trm{for}\ i \in \bb V, \quad \varphi_j(\tau = 0^-) = \vartheta_j \ \trm{for}\ j \in \bar{\bb V}.
\end{gathered}
}
The notation $\int^{(k)}$ means that the path integral variables within the region $\bb V$ on the $\tau = 0$ slice are constrained to equal the eigenvalues $k$ of central generators of $\A_{\bb V}$.\footnote{More precisely, we should also allow for constraints on links emanating from $\del \bb V$. Such constraints would play a role when computing, say, the entropy of the algebra $\G\_{D, \dot{D}}$ in eq.~\eqref{G D Ddot}. They will also play a role when studying various entropies of gauge theories. We will henceforth include this possibility when referring to boundary conditions at $\tau = 0$.}  The path integral is normalized so that
\bel{
  \sum_{k, \phi, \vartheta} \Psi_k^*(\phi, \vartheta)\Psi_k(\phi, \vartheta) = \sum_k \int^{(k)} [\d\varphi]\, \e^{-\int_{-\infty}^\infty L\, \d\tau} =  1.
}

\begin{figure}[tb!]
\begin{center}

\begin{tikzpicture}[scale = 1.5]




  \draw (-1.75, 0.025) -- (1.75, 0.025);
  \draw (-1.75, -0.025) -- (1.75, -0.025);

  \draw[thick, dotted] (-4, 0) -- (-2, 0);
  \draw[thick, dotted] (4, 0) -- (2, 0);

  \draw[red, very thick] (-2, 0) -- (-1.75, 0);
  \draw[red, very thick] (2, 0) -- (1.75, 0);
  \draw (-2, 0) node[red] {\scriptsize $\bullet$};
  \draw (2, 0) node[red] {\scriptsize $\bullet$};

  \draw[->] (3.5, 0.1) -- (3.5, 0.6);
  \draw (3.525, 0.45) node[anchor = west] {$\tau$};
  \draw (-3.75, 0.025) node[anchor = south] {$\tau = 0$};

  \filldraw[fill = white, white] (-2.75, -0.025) rectangle +(0.25, 0.05);
  \draw (-2.65, 0) node {$\vartheta$};
  \filldraw[fill = white, white] (2.7, -0.025) rectangle +(0.25, 0.05);
  \draw (2.85, 0) node {$\vartheta$};

  \draw (1.85, 0.05) node[anchor = south, red] {$k$};
  \draw (-1.85, 0.05) node[anchor = south, red] {$k$};

  \draw (0.1, 0.1) node[anchor = south] {$\phi^{(k)}$};
  \draw (-0.1, -0.1) node[anchor = north] {$\~\phi^{(k)}$};
\end{tikzpicture}

\end{center}
\caption{\small A side view of the path integral boundary conditions on the $\tau = 0$ slice, illustrating eq.\ \eqref{def rhoVk}. At all other values of $\tau$, the integral is unconstrained; even at $\tau = 0$, there is a sum over all values of fields $\vartheta$ in $\bar{\bb V}$, i.e.~outside the two red circles. The setup shown here calculates $\qmat{\~\phi^{(k)}}{\rho^{(k)}_{\bb V}}{\phi^{(k)}}$ when central generators are at or near $\del \bb V$; their locations are shown in red. The values of central generators are denoted $k$ and are the same at $\tau = 0^+$ and $\tau = 0^-$. The replica trick calculates $\Tr [\rho^{(k)}_{\bb V}]^n$ by taking $n$ copies of this setup and constraining $\phi^{(k)}$ on the replica $l$ to equal $\~\phi^{(k)}$ on replica $l + 1$, with $n + 1 \equiv 1$. The fields in the red region are then eigenstates of central generators with eigenvalue $k$ on each replica.}
\label{fig path int}
\end{figure}
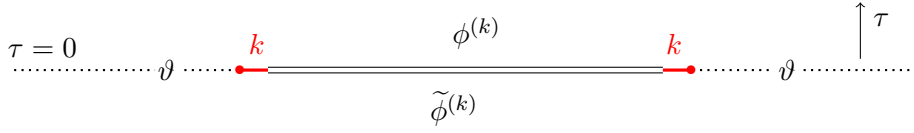

The reduced density matrix induced by the algebra $\A_{\bb V}$ is block-diagonal in the basis $\left\{\left\qvec{\phi^{(k)}\right}\right\}$, and each block has matrix elements
\bel{\label{def rhoVk}
\begin{gathered}
  \left\langle\~\phi^{(k)} \Big| \rho_{\bb V}^{(k)} \Big| \phi^{(k)} \right\rangle = \sum_{\vartheta} \Psi_k^*(\phi, \vartheta) \Psi_k(\~\phi, \vartheta) = \int^{(k)} [\d\varphi]\, \e^{-\int_{-\infty}^\infty L\, \d\tau}, \\
  \varphi_i(\tau = 0^+) = \phi^{(k)}_i, \quad \varphi_i(\tau = 0^-) = \~\phi^{(k)}_i \ \trm{for}\ i \in \bb V.
\end{gathered}
}
On the other hand, the normalized density matrix in sector $k$ (cf.~eq.~\eqref{def pk}) has matrix elements
\bel{\label{def bar rhoVk}
  \left\langle\~\phi^{(k)} \Big| \bar\rho_{\bb V}^{(k)} \Big| \phi^{(k)} \right\rangle = \int^{(k)} [\d\varphi]_k\, \e^{-\int_{-\infty}^\infty L \, \d \tau},
}
with the measure chosen so that $\Tr \bar \rho^{(k)}_{\bb V} = 1$ for all $k$. See fig.~\ref{fig path int}. For instance, for a spin chain with Dirichlet boundary conditions at the edge sites $\{v_1, v_2\}$ of the entangling interval $\bb V$, the normalization condition
\bel{\label{normalization k}
  \int^{(k)} [\d\varphi]_k\, \e^{-\int_{-\infty}^\infty L \, \d \tau} = 1
}
has $k = (k_{v_1}, k_{v_2}) \in \{\uparrow, \downarrow\} \times \{\uparrow, \downarrow\}$ labeling the four superselection sectors.

The sector weights $p_k$ appearing in \eqref{entropy p} can be calculated from the original path integral \eqref{def rhoVk}, as made clear by their definition \eqref{def pk}:
\bel{
  p_k = \Tr \rho^{(k)}_{\bb V} = \int^{(k)} [\d\varphi]\, \e^{-\int_{-\infty}^\infty L \, \d \tau}.
}
Note that here the measure of the path integral is the same as in \eqref{def Psi}, meaning that when summed over all $k$ the integrals in the above equation yield unity. The two measures are related by
\bel{\label{def measure}
  p_k\, [\d\varphi]_k = [\d \varphi].
}

These probabilities can also be obtained from the expectation values of central operators. Let $\{\O_i\}_{i = 1, \ldots, M}$ be the generators of the center of $\A_{\bb V}$, chosen so that their eigenvalues $k_i$ are roots of unity. A general operator in the basis of the central algebra has the form $\O^{(m)} \equiv \prod_{i = 1}^M \O_i^{m_i}$. Its expectation is
\bel{
  \avg{\O^{(m)}} = \sum_{k} p_k\, k_1^{m_1} \ldots k_M^{m_M} \equiv \sum_k \b K_{mk}\, p_k.
}
Note that the tuple $m$ is conjugate to $k$, and the matrix $\b K$ is simply the kernel of an $M$-dimensional Fourier transform.\footnote{For example, in a $\Z_2$ system, the eigenvalues $k_i$ of each central generator are $\pm 1$. They can be written as $\e^{\i \pi \kappa_i}$ for $\kappa_i \in \{0, 1\}$, and the expectations become $\langle\O^{(m)}\rangle = \sum_\kappa \e^{\i \pi \sum_i \kappa_i m_i} p_\kappa$. Note that $m_i = 0$ corresponds to the identity operator ($\O^{(0)} = \1$), which is always in the center. In subsection \ref{subsec edge mode action} we will show how this works for the case of a free compact scalar.} This ensures that $\b K$ is always invertible, and we can write
\bel{\label{p inversion}
  p_k = \sum_m \b K^{-1}_{km} \avg{\O^{(m)}}.
}
This expression can also be written as $p_k = \avg{\prod_i P_{k_i}}$, where $P_{k_i} \propto \sum_{m_i} k_i^{m_i} \O_i^{m_i}$ is the projector onto the $k_i$-eigenstate of the central generator $\O_i$.

Finally, let
\bel{\label{def Ik}
  p_k \equiv \e^{-I(k)}.
}
The sum over individual contributions of sectors to the entropy $S_{\bb V}$ can be written as
\algns{\label{S distillable early}
  S_{\bb V}\^{dist} \equiv \sum_k p_k S^{(k)}_{\bb V}
  &= - \pder{}n\Bigg|_{n = 1} \left[ \sum_k \e^{-I(k)} \int^{(k)} [\d \varphi]_k^n \, \e^{- \int_{(\Mbb \times \R)^n} L \, \d \tau}\right],
}
where $(\Mbb \times \R)^n$ denotes the $n$-replicated Euclidean spacetime, and $[\d\varphi]^n_k = [\d\varphi_1]_k \cdots [\d\varphi_n]_k$ is the replicated measure for the path integral in sector $k$, normalized according to eq.~\eqref{normalization k}. The same boundary condition $k$ holds on each copy of the region $\bb V$ on the $\tau = 0$ slice. The sum over boundary conditions (edge modes) does not depend on $n$ and can be commuted past the derivative $\pder{}n$.

The full entropy associated to the algebra $\A_{\bb V}$, given by eq.~\eqref{entropy p}, can be calculated by modifying the edge mode contribution in \eqref{S distillable} to get
\bel{
  S_{\bb V}
  = - \pder{}n\Bigg|_{n = 1} \left[ \sum_k \e^{- n I(k)} \int^{(k)} [\d \varphi]^n_k \, \e^{- \int_{(\Mbb \times \R)^n} L \, \d \tau}\right].
}
According to eq.~\eqref{def measure}, $\e^{-n I(k)} [\d\varphi]^n_k = [\d\varphi]^n$, so the full entropy can also be written as the path integral with the original measure on each replica, in agreement with eq.~\eqref{def S replica}:
\bel{\label{S full}
  S_{\bb V} = - \pder{}n\Bigg|_{n = 1} \left[ \sum_k \int^{(k)} [\d \varphi]^n \, \e^{- \int_{(\Mbb \times \R)^n} L \, \d \tau}\right].
}
We also record the expression \eqref{S distillable early} for $S\^{dist}_{\bb V}$ once eq.\ \eqref{def measure} is employed:
\algns{\label{S distillable}
  S_{\bb V}\^{dist} = - \pder{}n\Bigg|_{n = 1} \left[ \sum_k \int^{(k)} [\d \varphi]^n \, \e^{(n - 1) I(k) - \int_{(\Mbb \times \R)^n} L \, \d \tau}\right].
}

Let us clarify what these path integrals looks like in an Abelian gauge theory. Conventionally, path integrals are defined over variables $A_\ell$ on both spatial and temporal links. On spatial links, $\e^{\i A_\ell}$ are eigenvalues of $Z_\ell$, whereas --- as mentioned in footnote \ref{foot A0 role} --- $A_\ell$ on temporal links are Lagrange multipliers enforcing the gauge constraint. After imposing these constraints, we may gauge-fix the variables on temporal links to $A_\ell = 0$. We are left with time-independent gauge transformations; let us set them aside for a moment. In the axial gauge fixed so far, electric field eigenvalues are time derivatives $\del_\tau A_\ell$ of variables on spatial links. Eigenvalues of the electric central generators $E_v$ (defined in eq.~\eqref{def E abel}) are $\prod_{\ell \supset v, \, \ell \in \bb V}\exp\left\{\i\del_\tau A_\ell\right\}$  for each $v \in \del \bb V$. In the path integral \eqref{S full} associated to the maximal gauge-invariant algebra, within each $\int^{(k)}$ the fields are thus constrained to have
\bel{\label{electric center BCs}
   \e^{\i \sum_{\ell \supset v, \, \ell \in \bb V} \del_\tau A_\ell(\tau = 0)} = k_v, \quad v \in \del \bb V.
}
(The integral over temporal $A_\ell$'s emanating from $v \in \del \bb V$ makes sure that field configurations on $\bar{\bb V}$ end up having a matching flux $k$ at $\del \bar{\bb V}$.) The conditions \eqref{electric center BCs} are defined in axial gauge but are invariant under time-independent gauge transformations. Thus we can fix the remaining gauge freedom in any way we want, but we caution any prospective gauge-fixers that the replica structure makes it extremely natural to fix separately inside $\bb V$ and outside of it.

In nonabelian theories, the same story as above plays out, with one significant difference. The path integral \eqref{S full} for the maximal $\A_{\bb V}$ forces the gauge fields to have a definite representation of the electric flux through $\del \bb V$ for each sector $k$. Unlike in the Abelian case, an eigenstate with a definite flux representation is not invariant under time-indepenedent gauge transformations on $v \in \del \bb V$. (Recall from section \ref{sec subalgebras} that any state in $\H^R_{\bb V}$ has one ``dangling'' index $\alpha$ at each edge site, resulting in $\rho_{\bb V}$ having factors of $\dim R\times \dim R$ identity matrices.) We can \emph{choose} whether to fix the gauge on the edge; if we do not, these dangling degrees of freedom will get replicated and will contribute to $S_{\bb V}$. Gauge-fixing on $\del \bb V$ thus reduces $S_{\bb V}$ down to the ``gauge-invariant'' part $S\^{g\textrm{-}inv}_{\bb V}$ given by \eqref{def S ginv}. We will discuss the physical nature of this gauge-fixing in section \ref{s3}.

Equation \eqref{S full} should be taken as the regularized \emph{definition} of the replica trick path integral. It gives a precise treatment of boundary conditions and distinguishes between entropies associated to algebras with different centers. In general, the entropy associated to a general subalgebra $\A_{\bb V}$ can be calculated algorithmically by following these steps:
\begin{enumerate}
  \item Determine the center of $\A_{\bb V}$.
  \item Determine the set of degrees of freedom on which $\A_{\bb V}$ acts nontrivially, and pick the basis of the associated space $\H^*_{\bb V}$ so that the center generators are all diagonal. This sets the variables to be used in the path integral.
  \item Calculate the full entropy $S_{\bb V}$ via eq.~\eqref{S full}. In a nonabelian theory, the same equation can be used to calculate the gauge-invariant entropy $S_{\bb V}^{\textrm{g-inv}}$. It is obtained by fixing the time-dependent gauge transformations on $\del \bb V$.
\end{enumerate}
This procedure can be applied to any quantum theory, assuming we know how to discretize it such that we start off from a Hilbert space that factorizes.

\section{Miscellaneous comments} \label{sec comments}

\subsection{Gauge vs.~matter theories}

As repeatedly stressed throughout this note, there is no procedural difference between calculating entropies in matter and in gauge theories. The principal difference between these theories is that the latter always have central generators --- the Gauss operators. Moreover, in a gauge theory, the maximal gauge-invariant subalgebra supported on $\bb V$ always has additional central generators in the form of edge electric flux operators $E_v$, defined above eq.~\eqref{rho gauge}. In contrast, the maximal subalgebra on a region $\bb V$ has no center for matter theories.

In entropy calculations, Gauss operators in the interior of a region $\bb V$ do not give rise to any sums over superselection sectors. This is because all gauge-invariant states are in the same, singlet sector in which $G_v = \1$ holds as an operator equation. No other sectors are populated. This means that any reduced density matrix in $\bb V$ can be represented as a density matrix of gauge-invariant degrees of freedom times a projection to the singlet sector, eq.~\eqref{rho gauge}.

On the other hand, the existence of flux operators $E_v$ on $v \in \del \bb V$ means that gauge theories will naturally have nontrivial superselection sectors, and since these are supported on edges it is reasonable to refer to them as edge modes. (Here we assume that the ``natural'' algebra associated to a region is the maximal allowed algebra of operators on that region.) Using the ``balanced center'' procedure of \cite{Casini:2013rba} it is possible to find a non-maximal gauge-invariant algebra that has no edge modes. However, in a topological phase of a gauge theory this kind of algebra does not give rise to the usual topological entanglement entropy. In general, it is not known how universal data about the gauge theory are encoded in entropies of algebras with non-maximal centers.

\subsection{The edge mode action} \label{subsec edge mode action}

The general prescription in the previous section can be straightforwardly applied to $d = 1$ systems. However, $d > 1$ might seem hopeless, as the size of the center grows exponentially with $|\del \bb V|$. Fortunately, there are several situations in higher dimensions when the superselection sectors can be tamed. These are either free or infinitely gapped theories of various kinds, and here we will enumerate some examples. To simplify the discussion, we will focus on Dirichlet algebras for matter theories and on electric algebras for gauge theories. In both cases a sum over sectors generically appears.

\begin{enumerate}
  \item \emph{Product states}: The sum over superselection sectors is trivial if only one sector has nonzero weight. In such a case there is no ``classical'' entropy from the $-\sum_{k} p_k \log p_k$ term in \eqref{entropy p}. In a matter theory, this happens in massive phases with product states, such as in the Ising model at large external field. The ground state has all spins pointing in the same direction, say $Z_v = \1$. If the center of $\A_{\bb V}$ is generated by $Z$'s at the edge, only one superselection sector will be populated. Similarly, a lattice gauge theory at strong coupling will have a ground state with $X_\ell = \1$ on all links, and only one configuration of edge modes (electric fields piercing $\del \bb V$) will appear in the reduced density matrix.
  \item \emph{Topological phases}: All gauge theories with a finite gauge group have a topological phase at weak coupling.\footnote{If the gauge group is continuous, the topological phase may be inaccessible. Consider the case of $U(1)$.  We can imagine regulating it with $\Z_K$ for $K \gg 1$; the topological phase is reached when the coupling is much smaller than $1/K$ \cite{Radicevic:2015sza}, which is not a limit one usually considers in $U(1)$ theories.} In this limit, central generators are the electric flux operators $E_v$, and topological states have the same $p_k$ for each set $k$ of eigenvalues of $E_v$, modulo global constraints (as discussed in section \ref{sec algebras}). A quick way to see this is to note that all electric flux operators have zero expectation values, so $p_k = \frac1{K^{M - 1}}$ follows from eq.~\eqref{p inversion} (with $M$ denoting the number of sites at $\del \bb V$, and $\Z_K$ being the gauge group).
  \item \emph{Symmetry protection}: The previous two examples dealt with extreme weak and strong coupling limits of simple theories. It is conceivable that with a sufficient amount of symmetry,  only a few superselection sectors will be populated in more complicated models, as well. At the very least, we expect this to be the case in any symmetry-protected trivial (SPT) or symmetry-enhanced topological (SET) phases. We will see in section \ref{s3} that Chern-Simons theory naturally fits into this category, too. It would be interesting to discover other nontrivial theories in which the superselection sectors are simple due to a symmetry.
  \item \emph{Free theories}: In a free field theory the expectation values that enter eq.~\eqref{p inversion} are nontrivial, but they do factorize due to Wick's theorem. Consider a compact scalar, with central generators $Z_v \equiv \e^{\i \phi_v}$ for $v \in \del \bb V$. The expectations of central elements can be written as $\avg{\e^{\i \sum_v m_v \phi_v}} \sim \e^{-\sum_{v, u} G_{vu} m_v m_{u}}$. Plugging this back into \eqref{p inversion}, using $k_v \equiv \e^{\i \kappa_v}$ and $\b K^{-1}_{k m} \sim \e^{-\i \sum_v \kappa_v m_v}$, and summing over $m_v$, we find sector probabilities $p_k \sim \e^{-\sum_{v, u} G^{-1}_{vu} \kappa_v \kappa_{u}}$. This is an elementary way to derive the Gaussian distribution of edge modes that has played an important role in understanding the entanglement entropy in $d = 3$ Maxwell theory \cite{Donnelly:2014fua, Donnelly:2015hxa, Soni:2016ogt}.  Note that the edge mode action $I(k) = -\log p_k$ is nonlocal. The action that appears in the replica trick, eq.~\eqref{S distillable}, is thus not equivalent to the naive restriction of the original action to the subregion $\bb V$, as would be the case for an algebra $\A_{\bb V}$ without a center.

\end{enumerate}

\subsection{Lattice, continuum, and conformal boundary conditions}

The discussion so far was centered on path integrals carried out over lattice fields. When such integrals are dominated by configurations that vary over length scales much greater than the lattice spacing, lattice boundary conditions can be replaced by continuum ones. If the field configurations of interest vary slowly and the theory remains weakly coupled, the type of boundary condition (open, Dirichlet, Neumann, electric, magnetic, etc) will be the same in the continuum and on the lattice, although the edge mode action $I(k)$ may be renormalized. For instance, in a free theory we can approximate the sum over edge modes $k$ with an integral over boundary configurations $\varphi(x_\parallel)$ for $x_\parallel \in \del \bb V$, and these continuum edge modes will still have Gaussian weights, with an effective action $I[\varphi]$ \cite{Donnelly:2011hn}. This naive approach will not be possible for theories without an explicit Lagrangian description, for instance for strongly coupled theories defined as the infrared limits of known theories, where boundary conditions are more abstractly determined by bulk-to-boundary operator product expansions (see \cite{Liendo:2012hy} for a modern take on this). It is not known how to generalize our discussion to these cases, though of course if we were powerful enough to compute path integrals directly in the UV, our analysis would apply.

In arbitrary $d = 1$ CFTs, information about boundary conditions can be captured in a different, more efficient way. Conformal boundary conditions are classified by their transformations under the conformal group (with the space of boundary conditions spanned by Ishibashi states). Linear combinations of Ishibashi states, chosen in order to have nice modular properties, are called Cardy states \cite{Cardy:1989ir}. There is one Cardy state for each primary operator. In a study of the Ising CFT \cite{Ohmori:2014eia}, Cardy states at the entangling edges in the path integral construction were connected with choices of algebras.\footnote{Ref.~\cite{Ohmori:2014eia} focused on the R\'enyi entropies $(1 - n)^{-1} \log \Tr \rho_{\bb V}^n$ in the limit $n \rar \infty$. We here assume that the established connection between subalgebras and this particular limit of R\'enyi entropies can also extend all the way to $n = 1$, where the usual entanglement entropy is recovered.} In this reference it was also found that the leading universal term does not depend on these choices, which appears to be a non-generic phenomenon: as we will review in more detail below, the universal term depends on the choice of entropy in $d = 3$ free theories.

Specifically, for an interval $\bb V$, consider the maximal subalgebra $\A_{\bb V}$ of the Ising model $H = \sum_i (X_i X_{i + 1} + Z_i)$ and the corresponding Dirichlet-type algebras $\A\^{D, D}_{\bb V}$, $\A\^{D, O}_{\bb V}$, and $\A\^{O, D}_{\bb V}$ defined by dropping $Z$ generators from one or both of the edge sites. The algebra $\A_{\bb V}$ corresponds to open boundary conditions on the lattice. In terms of conformal boundary conditions, it corresponds to the Cardy states $\qvec\sigma$ on both entangling cuts. An open boundary condition on one edge site corresponds to the $\qvec\sigma$ state on that site in the CFT.

The Cardy states $\qvec\1$ and $\qvec\eps$ are more interesting, and they arise when Dirichlet conditions are imposed. The edge modes here are labeled by the pair $k = (k_1, k_2)$, $k_i \in \{\pm 1\}$, of eigenstates of edge operators $X_v$. When $k_1 = k_2$, the Cardy state is $\qvec{\1, \1}$ or $\qvec{\eps, \eps}$. (The two choices give the same path integrals, and they calculate the individual sector entropies $S^{(k)}_{\bb V}$.) When $k_1 = -k_2$, the Cardy state is $\qvec{\1, \eps}$ (or $\qvec{\eps, \1}$). When one boundary condition is open and the other Dirichlet, there are two superselection sectors labeled by, say, $k_1$, and the Cardy states on the edges are $\qvec{\1, \sigma}$ or $\qvec{\eps, \sigma}$, depending on the choice of $k_1$.

This is the only known example of a connection between conformal boundary conditions and algebras. It would be interesting to understand this link in more instances, and to check whether  the universal leading-order behavior in other $d = 1$ CFTs does not depend on the choice of algebra, as was found to be the case in the Ising model.

\subsection{Universal terms} \label{subsec universal terms}

In a CFT with a UV cutoff $\epsilon$ and in $d > 1$ space dimensions, it is common to say that the ground state entanglement entropy for a spherical region $\bb V$ of size $R$ takes the form
\bel{\label{univ}
  S_{\bb V} = a_{d - 1} \frac{R^{d - 1}}{\epsilon^{d - 1}} + a_{d - 3} \frac{R^{d - 3}}{\epsilon^{d - 3}} + \ldots + C_d \left\{
                                     \begin{array}{ll}
                                       -\gamma, & \hbox{$d$ even} \\
                                       a \log \frac R \epsilon, & \hbox{$d$ odd}
                                     \end{array}
                                   \right. + O(\epsilon/R).
}
Here $C_d$ is a theory-independent constant\footnote{Different authors have different conventions for this constant, and in section \ref{s3} we will simply absorb it into the definition of $a$ and $\gamma$.}, and $\gamma$ and $a$ are universal (regularization-independent) and can often be interpreted as counting the number of degrees of freedom in the CFT \cite{Ryu:2006ef, Nishioka:2018khk}. Which of the entropies defined so far should be understood when writing such formulas?

This question is not trivial; different entropy choices lead to different values of $\gamma$ and $a$. It is conceivable that they all have a counting interpretation and lead to $c$-theorems of interest \cite{Casini:2004bw, Casini:2012ei}. However, for odd $d$, the quantity $a$ can be connected to the trace anomaly of the CFT \cite{Solodukhin:2008dh}. (For even $d$, the trace anomaly vanishes and we do not have an analogous argument.) Briefly, the trace anomaly is calculated by the one-point function of the trace of the stress-energy tensor ${T_\mu}^\mu$ on a curved manifold, and the stress-energy tensor is defined as the variation of the action under a Weyl rescaling of the metric. In particular, if an entropy is computed by a replica trick path integral whose only dependence on the metric enters through the Lagrangian $L$ and the normalization of $[\d\varphi]^n$, then integrating over a replicated spacetime will yield a universal contribution proportional to the integral of $\avg{{T_\mu}^\mu}$ over the entanglement edge $\del \bb V$. This condition is fulfilled by the full entropy $S_{\bb V}$ in eq.~\eqref{S full}, but not by $S\^{dist}_{\bb V}$ or any of the $S^{(k)}_{\bb V}$'s, which all have additional nontrivial edge mode actions in the path integral. (Of course, in particular states the edge mode action may be metric-independent, as discussed in section \ref{subsec edge mode action}, but we have no reason to believe this happens in the ground state of any CFT.) On the other hand, the gauge-invariant entropy $S^{\textrm{g-inv}}_{\bb V}$ differs from $S_{\bb V}$ by gauge-fixing at every point along the entanglement edge; by spherical symmetry we expect this fixing to influence the area term and not the universal, subleading term. Thus, at this stage, we claim that only $S_{\bb V}$ and $S\^{g\textrm{-}inv}_{\bb V}$ can be entropies whose universal terms can generically match the trace anomaly.

We have also not specified \emph{which} algebra we want to compute the full or gauge-invariant entropy of. Based on calculations in free theories \cite{Donnelly:2014fua, Soni:2016ogt}, we conjecture that in all theories the trace anomaly matches the universal term in the entropy when $\A_{\bb V}$ is the \emph{maximal} algebra on region $\bb V$. Moreover, based on calculations in topological and confined phases of lattice theories \cite{Casini:2013rba, Radicevic:2015sza}, we also conjecture that the quantity $\gamma$ (sometimes also denoted $F$) that is customarily defined in even $d$ also matches the universal piece of the entropy of a maximal algebra on $\bb V$.

In the context of gauge theories, we will provide more plausibility arguments for this conjecture in section \ref{s3}. In particular, we will note that the topological entanglement entropy of Chern-Simons theory is reproduced only if no gauge-fixing is assumed at $\del \bb V$. Thus, we can sharpen our conjecture and claim that it is always the full entropy $S_{\bb V}$ of the maximal $\A_{\bb V}$ that contains the conventionally defined universal terms.

\subsection{Holographic entanglement entropy}

If a theory is holographic, the replica trick path integral can be calculated using a replicated bulk geometry, and the entropy associated to this bulk partition function is approximated by the area of the Ryu-Takayanagi surface in the classical limit \cite{Ryu:2006bv, Lewkowycz:2013nqa}. An important aspect of this story is that the Ryu-Takayanagi prescription does not involve a choice of subalgebra or of a type of boundary condition. Which of the many kinds of entropies that are associated to a region $\bb V$ does holography compute?

This question reduces to the one discussed in the previous subsection. The holographic entropy is known to reproduce the correct trace anomaly in its universal terms when the bulk contains Einstein gravity \cite{Ryu:2006ef}. Moreover, in higher derivative theories of gravity, the Ryu-Takayanagi prescription is modified so that the holographic entanglement entropy still computes the correct trace anomalies \cite{Hung:2011xb}. Building on the conjectures from the previous subsection, we propose that, in all dimensions, it is the full entropy of the maximal algebra on $\bb V$ that is computed by the holographic entanglement entropy, defined via Ryu-Takayanagi or its generalizations. It is not known what gravity constructions would calculate any of the other entropies that can be associated to $\bb V$.

\newcommand{\nn}{\nonumber}

\newcommand{\lb}{\left(}
\newcommand{\rb}{\right)}
\newcommand{\lcb}{\left\{}
\newcommand{\rcb}{\right\}}
\newcommand{\lsb}{\left[}
\newcommand{\rsb}{\right]}
\newcommand{\ld}{\left.}
\newcommand{\rd}{\right.}

\newcommand{\Sym}{\it Sym}

\renewcommand\be{\begin{equation}}
\newcommand\ba{\begin{eqnarray}}
\newcommand\ee{\end{equation}}
\newcommand\ea{\end{eqnarray}}

\newcommand\bone{{\bf 1}}

\definecolor{purple}{rgb}{0.7,0.0,0.5}
\definecolor{huh}{rgb}{0.0,0.6,0.8}
\definecolor{orange}{rgb}{1,0.5,0}
\definecolor{pink}{rgb}{1,0.4,0.4}
\definecolor{light-gray}{gray}{0.75}
\def\red#1{{\color{red} #1}}
\def\yellow#1{{\color{yellow} #1}}
\def\pink#1{{\color{pink}#1}}
\def\orange#1{{\color{orange}#1}}
\def\blue#1{{\color{blue}#1}}
\def\teal#1{{\color{huh}#1}}
\def\gray#1{{\color{light-gray} #1}}
\def\lc{\left\lceil}
\def\rc{\right\rceil}

\section{Local extensions of the Hilbert space in gauge theories}\label{s3}

In this section, we discuss the second universality class of approaches to define the entanglement entropy of a lattice gauge theory.
In contrast to the previous sections, here we will take the perspective that the Hilbert space of the gauge theory contains only the gauge-invariant states that faithfully represent the gauge-invariant operator algebra.
We then \emph{locally} extend this Hilbert space along the entangling cut by adding the minimal number of degrees of freedom at the cut so that it splits into tensor factors.
Although this operation may seem quite arbitrary, the two main points in this section are that {\it (i)} it is a physically meaningful procedure when the gauge theory emerges as an effective low-energy description of the physics, and {\it (ii)} it coincides with an especially natural set of boundary conditions in the replica trick, that roughly amount to smoothing out the conical singularity. (In section \ref{subsec universal terms} we arrived at an analogous requirement: that the replicated path integral depend on the metric only through the action and the usual measure.) Much of this section serves as a review of a cross-section of the literature on entanglement entropy in gauge theories in which these points were crystalized.

In cases where we can put the gauge theory on a lattice, the results obtained in this section coincide with the full entanglement entropy of the maximal algebra, \eqref{entropy p R}, in the conventions of the previous sections. This is because previously, we were already using the convention that the algebraic entanglement entropy for lattice gauge theories was defined w.r.t. a (globally) extended Hilbert space, as described in footnote \ref{f4} (and the extension of the Hilbert space away from the entangling surface does not affect the partial trace calculation).\footnote{In particular, earlier references to the ``full entropy of the maximal algebra" in a gauge theory include both the Shannon and $\log \dim R$ edge terms in the notation below.}

Organizationally, we will start by introducing the extended Hilbert space approach both on the lattice and in the continuum. We first do this at a formal level, then provide a physical justification for it. We then discuss the replica trick boundary conditions, and finally conclude with assorted comments.

\subsection{Definition}
\subsubsection{On the lattice}

The extended Hilbert space definition on the lattice is easiest to introduce by example. Consider Yang-Mills theory on a $d = 1$ spatial lattice with two sites connected by two links, and suppose that we want to assign a reduced density matrix to one of the links. This is the problem of computing the entanglement across an interval in $d = 1$ Yang-Mills theory on a spatial circle \cite{Donnelly:2014gva}.

Yang-Mills in two spacetime dimensions is solvable, see e.g.\ \cite{Cordes:1994fc}. It's known for instance that the Hilbert space on the circle is the space of square-integrable class functions on the gauge group $G$ ($L^2$-functions on the group manifold s.t.\ $\psi(gug^{-1}) = \psi(u)$ for all $g \in G$), for which a convenient basis is furnished by the characters,
\be\label{hphys}
\H\_{phys} = \mbox{class functions in } L^2(G) = \trm{span}\{|R\rangle\}\,.
\ee
The gauge-invariant operators in the theory are the Casimirs built out of the electric fields, which are diagonal in the representation basis, and the Wilson loop operators in all the representations of $G$.
The Hilbert space \eqref{hphys} doesn't refer to the underlying spatial manifold and so clearly doesn't factorize. In the extended Hilbert space prescription, we're instructed to embed it into the minimal larger one that does.

Intuitively, the presence of extended operators that cannot be split into local constituents (i.e.\ the Wilson loops) is what prevents the Hilbert space that faithfully represents the operator algebra from factorizing. We can get around this by adding pairs of non-dynamical (infinitely massive) surface charges in all the representations of the gauge group at the ends of our entangling interval, which allow us to cut the Wilson loops into Wilson lines.
This is tantamount to lifting the Gauss constraint at the lattice sites, so the extended Hilbert space is the tensor product of two link Hilbert spaces, which are spanned by the matrix elements of the group representations:
\be
\H\_{ext.} = \H\_{left} \otimes \H\_{right}\,, \qquad \H\_{left/right} =  L^2(G) = \trm{span}\{|R_{\alpha\beta}\rangle\_{left/right}\}\,.
\ee
(See section \ref{subsec nonabelian gauge theories} for more details.) Under a gauge transformation $\Lambda \in G$ on the first site, these states transform as
\bel{
  \qvec{R_{\alpha\beta}}\_{left} \mapsto \sum_{\alpha'} R^{-1}_{\alpha \alpha'} (\Lambda) \qvec{R_{\alpha' \beta}}\_{left},\quad \qvec{R_{\alpha\beta}}\_{right} \mapsto \sum_{\alpha'} R_{\alpha\alpha'}(\Lambda) \qvec{R_{\alpha' \beta}}\_{right},
}
and under a gauge transformation on the second site they transform as
\bel{
  \qvec{R_{\alpha\beta}}\_{left} \mapsto \sum_{\beta'} R_{\beta\beta'}(\Lambda) \qvec{R_{\alpha \beta'}}\_{left},\quad \qvec{R_{\alpha\beta}}\_{right} \mapsto \sum_{\beta'} R^{-1}_{\beta\beta'}(\Lambda) \qvec{R_{\alpha\beta'}}\_{right}.
}

The physical Hilbert space $\H\_{phys}$ is embedded into $\H\_{ext.}$ as
\be\label{rij}
|R\rangle \rightarrow \frac1{\dim R} \sum_{\alpha,\,\beta \in 1,\, \dots,\, \dim R} |R_{\alpha\beta}\rangle\_{left} \otimes |R_{\beta\alpha}\rangle\_{right}.
\ee
It is easy to see that the state above is indeed gauge-invariant. The physical picture is that, with the matrix indices labeling the infinitely massive surface charges, we want pairs of charges across the entangling cut to transform oppositely under gauge transformations in order to make a gauge-invariant state, so we assign them the same index.

We can now take the most general state $|\Psi\rangle = \sum_R \psi_R |R\rangle \in \H\_{phys}$, embed it in the extended Hilbert space as $\sum_{R_{\alpha\beta}} \frac{\psi_R}{\dim R} |R_{\alpha\beta}\rangle\_{left} \otimes |R_{\beta\alpha}\rangle\_{right}$,
and compute the reduced density matrix by tracing out, say, the right  tensor factor in the extended Hilbert space. Its von Neumann entropy is
\be\label{2dym}
S\_{left} = -\sum_R p_R \log p_R + 2\sum_R p_R \log \dim R
\ee
where $p_R = |\psi_R|^2$\,. The first term is a probability distribution over the states; we'll call this the Shannon entropy. The second, $``\log \dim R"$ term, comes from the statistics of fusing the non-dynamical surface charges to make a singlet. (The factor of 2 comes from the entangling interval having two ends.)
This concludes our review of the $d = 2$ example.

In higher-dimensional lattice gauge theories, the extended Hilbert space idea was discussed in refs.\ \cite{Buividovich:2008gq, Donnelly:2011hn, Ghosh:2015iwa}.\footnote{The conventions in these references differ by some cosmetic details. In \cite{Donnelly:2011hn}, the entangling cut intersects a collection of boundary links instead of boundary sites, and one extends the Hilbert space by adding a new lattice site at each such intersection where one doesn't impose the Gauss law. In \cite{Ghosh:2015iwa}, the extended Hilbert space on the lattice is taken to be the ``globally extended" one, $\H_{\bb V} = \otimes\H_\ell$, instead of just lifting the Gauss law at the boundary sites. But as explained in section \ref{sec subalgebras}, an extension of the Hilbert space away from the entangling cut doesn't affect the entanglement entropy of states in the gauge-invariant subspace.}
There we find the same types of edge terms as in \eqref{2dym} when we promote the $R$'s that label the edge states $\{|R\rangle\}$ to instead label superselection sectors specified by the eigenvalues of the Casimirs at all the boundary lattice sites (as mentioned in section \ref{sec subalgebras}, where these were included in the general sector labels $k$). Then the entropy associated to a region $\bb V$ is
\be\label{lgtee}
S_{\bb V} = -\sum_R p_R \log p_R + \sum_R p_R \log \dim R - \sum_R p_R \Tr (\rho_R \log \rho_R)\,,
\ee
$\rho_R$ being the normalized density matrix in the superselection sector $R$ (see eq.\ \eqref{def SVk}). The first two terms are the edge terms from each lattice site along the boundary, and the third captures the distillable \cite{Ghosh:2015iwa} entanglement of interior degrees of freedom, that the $d = 2$ example was too simple to support.

In short, we see that the extended Hilbert space prescription on the lattice agrees with the full entanglement entropy, \eqref{entropy p R}, w.r.t.~the globally extended conventions of the previous sections. This was first pointed out in \cite{Ghosh:2015iwa}.

It's interesting to compare this result to the entropy $S_{\bb V}\^{g\textrm{-}inv}$ of the maximal gauge-invariant subalgebra/electric algebra {\it with respect to the gauge-invariant Hilbert space}.
As discussed around \eqref{entropy p R}, \eqref{def S ginv},
\be\label{S alg ginv}
S_{\bb V} = S_{\bb V}\^{g\textrm{-}inv} + \sum_R p_R \log \dim\,R\,;
\ee
the entropy of the maximal gauge-invariant subalgebra in the region ${\bb V}$ contains the Shannon edge term and distillable entanglement in \eqref{lgtee}, but not the $\log \dim\, R$ edge term. From the point of view of the infinitely massive charges that we introduced in this section, this isn't surprising. The edge term basically measures the correlations of the charges, that aren't part of our gauge-invariant operator algebra.

Note that the expression ``$\log \dim R"$ is the expectation value of an IR operator which is a function of the gauge group Casimirs. Hence, the result from the extended Hilbert space prescription is gauge-invariant despite that we got it from carrying around gauge-variant data.

\subsubsection{In the continuum}

As mentioned above, one way to define the extended Hilbert space on the lattice is to lift the Gauss constraint at boundary lattice sites. The naively analogous operation in the continuum would be to quantize the phase space after partial gauge-fixing in the time direction (i.e.\ $A_0 = 0$), but without then imposing the residual gauge symmetry at the entangling surface $\partial \bb V$, instead promoting the time-independent gauge symmetry to a global symmetry along the entangling surface. (See the discussion below eq.~\eqref{S distillable}.) The embedding $\H\_{phys} \subset \H\_{ext.}$ is then fixed by demanding that the states in $\H\_{phys}$ are singlets w.r.t.\ this new global symmetry.

We will say more about the formal justification for this operation below, but for now let us put it to use in an example.
By following these steps we can recover the topological entanglement entropy (TEE) in Chern-Simons (CS) theory in $d = 2$, as recently discussed in \cite{Fliss:2017wop, Wong:2017pdm}.
Consider the CS theory on a spacetime $\Mbb \times \R$.
Its action is
\be\label{csa}
S\_{CS} = \frac k {4\pi}\int_{\Mbb \times \R} \Tr\left(A \wedge dA + \frac 23 A \wedge A \wedge A\right)\,.
\ee
On a compact $\Mbb$, its Hilbert space is the space of conformal blocks of the chiral WZW model on $\Mbb$ with gauge group $G$ and level $k$ \cite{Witten:1988hf, Elitzur:1989nr}. In particular, the Hilbert space of the CS theory on $\Mbb = {\bf S}^2$ is one-dimensional and can't be meaningfully partitioned. On the other hand, the TEE of the Abelian CS theory across a disk $\bb V \subset \Mbb$ is \cite{Kitaev:2005dm}
\be
\gamma\_{CS} = \frac 12 \log k.
\ee
This is the universal piece of the entanglement entropy $S_{\bb V}$ (see eq.~\eqref{univ}), which is clearly nonzero for $k > 1$. This raises the question of what it is that we are counting.

We can answer this by looking at the canonical quantization of CS theory on a disk, $\Mbb = {\bf D}^2$ \cite{Elitzur:1989nr}.
Let us briefly review the steps in that paper.
We first chose the gauge $A_0 = 0$, so that the action becomes
\be\label{scs1}
S\_{CS} = \frac k{4\pi}\int_{\Mbb \times \R}\, \Tr(\tilde A\, \partial_t\tilde A) \,,
\ee
where the tilde means that we are restricting to the space components. We should also keep track of the Gauss law $F_{12} = 0$ and the residual time-independent gauge symmetry.  The Gauss law can be solved on the disk with $\tilde{A} = \tilde\d U U^{-1}$, for $\tilde\d$ the spatial exterior derivative, and $U \in G$. Plugging this into \eqref{scs1}, we find
\be\label{lwzw}
S\_{CS} = S\_{WZW} = \frac k{4\pi} \int_{\partial \Mbb} \Tr(U^{-1}\partial_\phi U U^{-1}\partial_t U) \d\phi\, \d t + \frac{k}{12\pi}\int_{\Mbb} \Tr(U^{-1}\d U)^3\,,
\ee
where $\phi$ is the angular coordinate on $\partial \Mbb$. This is the chiral WZW action.
At this point, the authors of \cite{Elitzur:1989nr} conclude that the CS Hilbert space on the disk is the Hilbert space of the WZW theory on the boundary.

According to the bookkeeping so far, we seem to have decided by fiat to forget about the residual time-independent gauge symmetry at the boundary of the disk, promoting it to be a global symmetry instead. Historically, the argument was that the action \eqref{csa} is not gauge-invariant in the presence of a boundary, and indeed it was observed early on that this choice is needed to consistently glue Euclidean path integrals along open cuts (i.e.\ to get $Z_M = \int [d\psi] Z_{(M_1,\Sigma)}[\psi] Z_{M_2, \Sigma}[\psi]$ when we cut a closed manifold $M$ into $M_1, M_2$ along $\Sigma$) \cite{Witten:1991mm}.
In our present context though, this construction of $\H_{\bf D^2}$ allows us to utilize the extended Hilbert space prescription, with
\be
\H\_{ext.} = \H_{{\bf D}^2} \otimes \H_{{\bf D}^2}\,.
\ee

To finish computing the entanglement entropy of the disk $\bb V$, all that remains is for us to decide how the unique state on ${\bf S}^2$ is embedded in the Hilbert space of the two disks. The defining property is that it should transform oppositely under the global symmetries of the two WZW CFT's in order to form a singlet, as the continuum analog of pairing up the matrix indices in \eqref{rij}. This uniquely specifies the embedding of the state on ${\bf S}^2$ in $\H_{\rm ext.}$ to be
%
\be\label{ishs}
|I \rangle = \sum_n |\1, \bar n \rangle \otimes |\1, n\rangle\,,
\ee
 where the sum runs over the (infinitely many) descendant states in the conformal module of the primary field associated with the identity representation of $G$.

 Finally, tracing out one of the tensor factors in \eqref{ishs} with a thermal regulator $e^{-\epsilon H}$ \cite{Wen:2016snr, Das:2015oha}, we find
\be
S\_{CS} = \left(\mbox{non-universal terms}\right) - \frac 12\log k\,.
\ee
In this way we see that the prescription described above is able to correctly compute the TEE.%
\footnote{
One somewhat unsatisfactory aspect of this calculation is that the factor of $k$ was buried in the thermal regulator, so although we are counting edge modes of the topological medium, we can't point to $``\sqrt{k}$ microscopic entangled degrees of freedom". A related situation where we can be somewhat more precise  if we quantize Chern-Simons theory on ${\bf S}^2$ with a Wilson line ending on static non-dynamical charges. In this case, repeating the above steps, we would find eq.  \eqref{ishs} for the analogous state corresponding to the representation assigned to the Wilson line, and a TEE differing from the identity representation by $\log \dim R$. This is the $\log \dim R$ type edge term of the Wilson line, quantifying the maximal entanglement of the static charges. }

The origin of the continuum extended Hilbert space prescription can be understood more rigorously as follows.\footnote{We thank Ronak Soni for discussions on these  points.}
For a gauge theory on a space with a boundary, requiring the boundary part of the variation of the action to vanish usually leads to a {\it set} of allowed types of boundary conditions that we can pick from to define the theory on that space.

For instance, consider the generator of time-independent gauge transformations $Q = \int_{\Mbb} \vec E \cdot \vec\nabla\lambda$ in a Maxwell theory, with $\lambda$ the gauge parameter. (The generator for time-dependent gauge transformations doesn't have a boundary term on a manifold with spatial boundaries.) Integrating by parts, we see that on a closed manifold this operator vanishes by the Gauss law, but in the presence of a boundary,
\be\label{qbdy}
Q = \int_{\partial\Mbb} \lambda\,  E_\perp,
\ee
where $E_\perp$ is the component of the electric field perpendicular to $\partial\Mbb$. As long as $Q$ isn't set to zero by the choice of boundary conditions,
it will act nontrivially on the states satisfying the Gauss law,
and can be seen as the generator of a large gauge transformation, i.e., a global symmetry that acts on the boundary. (More precisely, in this case the boundary gauge transformations are not zero modes of the symplectic form \cite{Donnelly:2016auv}.) In the Maxwell theory, we can set $Q = 0$ by choosing the electric conditions $E_\perp = 0$.

In Chern-Simons theory, like in any other theory with symmetry-protected edge states, there is no choice of boundary conditions which can kill off its version of $Q$. A straightforward way to see this is to consider the gauge variation of the Abelian Chern-Simons action in the presence of the boundary,
\bel{
  S\_{CS} \mapsto S\_{CS} + \frac k{4\pi} \int_{\del \Mbb \times \R} \lambda\, E_\parallel.
}
If the gauge constraint isn't relaxed at the boundary, the electric fields along it will have to vanish to make $Q = 0$, but this is impossible because the CS action requires there to be no magnetic flux, even right by the boundary, and it is impossible to enforce both $E_\parallel = 0$ and $B = 0$ at $\del\Mbb$, as these variables don't commute.

This analysis shows that both Chern-Simons and Maxwell theory admit consistent boundary conditions in which $A_0|_{\partial \Mbb} = 0$ is fixed using time-dependent gauge transformations, no additional constraint is imposed on the gauge-invariant fields at $\tau = 0$\footnote{Technically, this is true if we take the subregion to be a causal diamond instead of $\mathbb{V} \times \mathbb{R}$. E.g.\ for Maxwell theory, the boundary conditions that lead to nontrivial large gauge transformations are $A_0|_{\partial \Mbb} = 0$ and $B_\parallel = 0$ on $\mathbb{V} \times \mathbb{R}$. The latter constrains a gauge-invariant operator so these BC's do not implement the extended Hilbert space prescription. But on a causal diamond, the second BC becomes $\rho B_\parallel$ for $\rho$ the local Rindler radius, that vanishes on $\partial\mathbb{V}$ \cite{Blommaert:2018rsf}.}, and the appropriate generator $Q$ of time-independent gauge transformations is nonvanishing on the boundary.
Choosing these boundary conditions on $\partial\Mbb$ and its complement allows us to construct the continuum extended Hilbert space with $Q$ having a nontrivial action on the edge modes.


\subsection{Path integral boundary conditions}

In section \ref{sec path integrals}, we explained how the entanglement calculation for different choices of gauge-invariant subalgebras on the lattice lead to different boundary conditions in the replica trick path integral.
Which path integral boundary conditions does the extended Hilbert space prescription correspond to, even when applied to theories with no known lattice regulator? This question turns out to have the following nice answer in a wide range of compact gauge theories: {\it the extended Hilbert space prescription corresponds to open boundary conditions in the replica trick}.

For theories that we can discretize, this fact can be understood from the perspective of the earlier sections as follows. On the lattice, the extended Hilbert space definition equals the entropy of an extended, gauge-variant operator algebra that includes all the open Wilson lines ending on the entangling surface (see \cite{Ghosh:2015iwa, Soni:2016ogt} for a related discussion). This larger algebra has no center (discounting the Gauss operators in the interior), so the path integral should have no constraints.

Several instances of this general claim (including also in Chern-Simons theories that we don't know how to discretize), were discussed piecewise in the literature. We finish this section by listing them here.

\subsubsection{The replica trick in solvable gauge theories}

In the Hartle-Hawking states of $d = 1$ Yang-Mills theory  \cite{Donnelly:2014gva, Gromov:2014kia} and $d = 2$ CS theory \cite{Dong:2008ft} that are set up by a Euclidean path integral on a half-sphere, for any choice of entangling region, {suppose we assume that the conical defect in the $n$-replicated manifold can be smoothed out without changing the topology of the manifold.} E.g.\ for CS theory on a spatial ${\bf S}^2$, suppose that the $n$-replicated manifold for the entanglement entropy of a region $\bb V$ with disk topology has the topology of a smooth ${\bf S}^3$.
Then the replica trick becomes quite straightforward to implement. E.g.\ for the CS theory,
\be\label{csee}
\frac{Z_n}{Z_1^n} = \frac{Z(\b S^3)}{Z(\b S^3)^n} = (\S^0_0)^{1-n} \qquad \implies \qquad S_{\bb V} = -\partial_n \left. \frac{Z_n}{Z_1^n}\right|_{n=1} = \log \S_0^0\,,
\ee
where we made use of formulas relating the CS partition function on three-manifolds to matrix elements of the modular S-matrix \cite{Witten:1988hc}, e.g., $Z(\b S^3) = \S_0^0\,.$ \footnote{See \cite{Schnitzer:2016lrd}, \cite{Schnitzer:2016xaj} for replica trick calculations in the dual WZW theories.}

A natural question is which Lorentzian prescription this ``naive replica trick" corresponds to.
In the examples that were studied in the literature, the answer is the extended Hilbert space prescription.
The CS calculation \eqref{csee} is one example, since both the ``naive replica trick" and the extended Hilbert space prescription correctly reproduce the TEE.
Also in $d = 1$ Yang-Mills theory, the ``naive replica trick" includes the $\log \dim\,R$-type edge term, \eqref{2dym}, as an empirical fact (see ref.~ \cite{Donnelly:2014gva} for the calculation).
In cases where the gauge theory can be put on a lattice (which is clear for Yang-Mills but not necessarily Chern-Simons theory), this result is merely a manifestation of our general statement above, where the extended Hilbert space corresponds to open boundary conditions at the entangling cut that then can be contracted to a point in the continuum limit.
See \cite{JafferisIfQ, Cardy:2016fqc, Donnelly:2016jet} for related comments.


\subsubsection{Conformal anomaly for a $U(1)$ gauge theory}

In a continuum QFT, the entanglement entropy is dominated by UV-sensitive divergent terms. However, in even (odd) spacetime dimensions, the coefficient of the log-divergent (constant) term in an expansion in powers of the UV cutoff is expected to be universal, \eqref{univ}.

For a Maxwell theory in $d = 3$, the computation of the log coefficient in the entanglement entropy of a ball-shaped region was historically the subject of some dispute. This theory is conformal. In CFTs, the log coefficient of the entanglement entropy of a ball-shaped region is a (known) function of the central charges and the extrinsic curvature of the entangling boundary on symmetry grounds. In a $d = 3$ CFT, this universal term is proportional to the $a$-anomaly  \cite{Ryu:2006ef}, and equals $\frac{31}{45}\log(R/\epsilon)$ (with $R$ a length scale associated to the entangling region and $\epsilon$ the UV cutoff) for the $U(1)$ theory in particular.

However, a different answer was found in the literature \cite{Dowker:2010bu} using what is now called the CHM map \cite{Casini:2011kv}.
A general method to compute the entropy a ball-shaped region of a CFT, that relies on just the conformal symmetry, is to map the domain of dependence of the ball-shaped region to a hyperbolic or de Sitter spacetime, and the entanglement entropy to the thermal entropy on this space \cite{Casini:2011kv}. One can then find the thermal entropy by integrating the energy density with $\d E = T \d S$, or by finding the partition function on the dS/hyperbolic space with a periodic time $\beta$, and using the geometric entropy formula%
\footnote{%
Note that in the CHM map to hyperbolic space, $Z(\beta)$ refers to the partition function on ${\bf H}^{d-1} \times {\bf S}^1_\beta$ which is a smooth space without a conical singularity for all $\beta$. On the other hand, if we are doing the replica trick on a space where we use the geometry to provide the analytic continuation \cite{Callan:1994py}, as e.g. when computing the entropy in the Rindler wedge (section \ref{s723}), $Z(\beta)$ refers to the partition function on a manifold which is singular for $\beta \neq 2\pi$.
}
\be\label{geome}
S = (1-\beta\partial_\beta)\log Z(\beta)|_{\beta = 2\pi}\,.
\ee
From the dS energy density,
Dowker \cite{Dowker:2010bu} found a different answer, $a_{\rm eff} = 16/45$ (see also ref.\ \cite{Casini:2015dsg}).

The hyperbolic analog of this discrepancy was clearly explained in \cite{Huang:2014pfa} (see also refs.\ \cite{Eling:2013aqa} for the analogous explanation in dS space, which requires treating a conical singularity and is similar to section \ref{s723} below, as well as \cite{Soni:2016ogt}, which showed by directly analyzing the entanglement entropy of $U(1)$ gauge theory on the lattice that the Shannon edge term contributes the difference of $1/3$).
The upshot is that the CHM prescription by itself is incomplete: in the thermal entropy problem, one also has to pick boundary conditions as the analog of the boundary conditions at the entangling surface, and different choices give different answers.
In particular, if we replace the entanglement entropy problem with the problem of computing the partition function of a $U(1)$ gauge theory in hyperbolic space, we have to pick boundary conditions at the asymptotic boundary. Computing the thermal entropy with implicitly fixed asymptotic boundary conditions in a heat kernel approach \cite{Huang:2014pfa}, one finds $a_{\rm eff} = 16/45$.  However, one can account for the discrepancy by adding to the heat kernel calculation the partition function of a free scalar on ${\bf S}^2$, which is the amount of residual gauge symmetry on the asymptotic boundary after we fixed the time-dependent gauge symmetry.
Hence, the extended Hilbert space prescription gives the correct conformal anomaly, and the difference of $1/3$ comes from the Shannon edge term.

Incidentally, this discussion is an explicit example of the principle that the choice of prescription for computing the entropy affects its universal term; see section \ref{sec path integrals}.\footnote{Relatedly, in case the Maxwell theory emerges from massive but dynamical charges, one finds a third value for the log coefficient, $a_{\rm eff} = \frac{16}{45}+1$ \cite{Pretko:2015zva}, \cite{Pretko:2018yxl}. In this case, the dynamical charges are screened with a thermal entropy in a layer of size $m^{-1}$ around the entangling boundary.}


\subsubsection{Kabat contact term in the geometric entropy}\label{s723}

For $d = 3$ Maxwell theory, another longstanding and closely related puzzle in the literature was the following. Consider the entanglement entropy of the half-space (Rindler wedge) $\bb V$ in the vacuum of the Maxwell theory. If we apply the geometric entropy formula \eqref{geome} to compute the entanglement entropy, where $Z(\beta)$ is the partition function on a manifold with a conical singularity at the origin for $\beta \neq 2\pi$, the entropy will acquire a contact term with the conical singularity \cite{Kabat:1995eq}, basically because the the spin-1 Laplacian is $\square \delta_{\mu\nu} + R_{\mu\nu}$. For twenty years it was an open problem if this ``Kabat contact term'' had a state-counting interpretation.

This was recently answered in the affirmative by Donnelly and Wall \cite{Donnelly:2014fua, Donnelly:2015hxa} who explicitly showed the following: Kabat's $Z(\beta)$ is equal to the partition function $Z\_{bulk}$ on the replicated space with a tubular region removed around the conical defect at $\partial \bb V$ and the boundary condition $E_\perp = 0$ placed there,
multiplied by
\be\label{zedge}
Z\_{edge} = \int [\d E_\perp]\, e^{-I(E_\perp)}\,,
\ee
where $I(E_\perp)$ is the on-shell Euclidean action of the edge modes (the continuum version of $I(k)$ from eq.~\eqref{def Ik}), and the measure is obtained by taking the continuum limit of the discrete lattice measure. In short, we impose open boundary conditions around the defect. Here we note that we can use $E_\perp = 0$ to compute the entropy in each superselection sector; the distillable entropy is independent of $E_\perp$, since the theory is free. Thus the path integral factorizes into $Z = Z\_{bulk}Z\_{edge}$. Hence, the statistical interpretation of the Kabat contact term is that it counts the Shannon edge modes of the $U(1)$ gauge theory.


\subsection{Short remarks}

\subsubsection{Emergence}

Until now, our discussion of the extended Hilbert space was formal, and perhaps in poor taste from the usual point of view that gauge symmetries are fictional.
A physical justification for it is that if we replace the extended Hilbert space with a local UV Hilbert space in an emergent gauge theory, then \eqref{lgtee} will hold up to a constant that is independent of the state in the IR. One situation where this is trivially true is the toric code and ``globally extended Hilbert space" of section \ref{sec subalgebras}.

Ref.\ \cite{Lin:2017uzr} argued that it is true more generally.
The argument is the following.
On the one hand, in a low-energy emergent gauge theory, the UV reconstructions of IR Wilson loops as composite operators are able to factorize along $\partial \bb V$ by assumption.
On the other hand,
if we think of the entanglement entropy as an entropy of the maximal UV operator algebra on the region, the explicit expansion of the density matrix, eq.~\eqref{def rho}, has zero support on UV operators that are not either the UV analogs of gauge-invariant operators or of Wilson lines ending on $\partial \bb V$. So the UV-exact entanglement entropy can only differ from the entropy of the ``IR gauge-invariant algebra" by a constant related to the relative sizes of the Hilbert spaces in the UV and the IR.\footnote{This constant arises from the different definitions of the trace used in calculating the von Neumann entropy, as discussed around eq.~\eqref{S alg ginv}.}

\subsubsection{Duality}

An open question is how the edge terms map under duality. Euclidean replica trick results for (Seiberg-)dual nonabelian gauge theories appear to agree \cite{Nishioka:2013haa}.
This result can be decomposed into the sum of distillable entropy and the entropy coming from edge terms.
Given an exact duality map for all operators and states, the algebraically defined entropies $S\^{g\textrm{-}inv}_{\bb V}$ of dual algebras must trivially agree, but the $``\log \dim R"$ edge terms are more mysterious, since they seem to depend explicitly on the gauge group. Do the $``\log \dim R"$-type edge terms dualize to edge terms nonetheless? Preliminary evidence suggests yes \cite{Lewkowycz:2013laa}, but it would be nice to clarify this in an analysis along the lines of ref.\ \cite{Donnelly:2016mlc} for Abelian gauge theories.


\section{Conclusion} \label{sec conclusion}

We have given precise path integral prescriptions for calculating various entropies associated to a given subalgebra $\A_{\bb V}$. Our approach is elementary and applies to a wide array of theories, and we have shown how to connect it to various other recipes for calculating entropy in QFTs. While the results of this paper do not facilitate the computation of particular replica trick path integrals, they do show how they should be understood in any formal proof concerning properties of entanglement entropy. Several interesting questions remain open, however.

As emphasized in section \ref{sec entropies}, several different yet natural types of entanglement entropy exist for each choice of algebra on a given region. We have seen that certain special prescriptions for calculating entropy (holography and CHM-like maps in conformal theories with appropriate boundary conditions) correspond to the full entropy of the maximal algebra $\A_{\bb V}$ on $\bb V$ in a small set of theories (holographic theories with Einstein gravity duals, the Ising CFT, free theories). An immediate goal is to fill in the blanks as much as possible. For instance, what bulk quantities calculate the full entropy of a non-maximal algebra? Answers to such questions may teach us more about the bulk reconstruction of operators.

Another set of notions that we have not studied involves extending our discussion to other measures of entanglement, such as mutual information, entanglement negativity, covariant entanglement entropies, etc. In particular, it would be interesting to find the analogues of the various choices directly in the Tomita-Takesaki framework for continuum QFTs \cite{Witten:2018zxz}. An important ingredient in any analysis of the monotonicity of entropic measures will be a careful definition of which representation of $\A_{\bb V}$ is appropriate for a given type of center.

Finally, we remark that it may be possible to associate entropies (or other measures of entanglement) to sets of operators that are not necessarily algebras \cite{Ghosh:2017gtw}. This provides yet another direction in which our analysis of centers and boundary conditions could be generalized.

\section*{Acknowledgements}

We thank Rob Myers, Tatsuma Nishioka, and Kantaro Ohmori for useful conversations, and Lorenzo Di Pietro and Ronak Soni for discussions and comments on a draft of this paper. JL is supported by the William D. Loughlin Membership at the IAS and by the U.S. Department of Energy, and would also like to thank the Perimeter Institute and Galileo Galilei Institute for hospitality as this work was being completed. \DJ R thanks the Galileo Galilei Institute and the ACRI fellowship, the University of Tokyo, and the Kavli Institute for the Physics and Mathematics of the Universe for hospitality and support while this work was in progress. Research at Perimeter Institute is supported by the Government of Canada through Industry Canada and by the Province of Ontario through the Ministry of Economic Development \& Innovation.

\bibliographystyle{ssg}
\bibliography{review2}

\end{document}